\documentclass[epj-spec]{svjour}
\usepackage{graphicx}
\usepackage{psfrag}
\usepackage{amsmath}
\usepackage{amsfonts}
\usepackage{relsize}
\usepackage{hyperref}
\usepackage[numbers]{natbib}
\usepackage{slashed}

\DeclareMathOperator{\Tr}{Tr}
\makeatletter
\def\makeheadbox{{%
\hbox to0pt{\vbox{\baselineskip=10dd\hrule\hbox
to\hsize{\vrule\kern3pt\vbox{\kern3pt
\hbox{{\bfseries Manuscript submitted to \@journalname.}\hspace{6em}
  SPIN-07/23, ITP-UU-07/34, DCPT-07/39}
\hbox{(to appear in the proceedings of the 45th Winter School on
  Theoretical Physics)}
\kern3pt}\hfil\kern3pt\vrule}\hrule}%
\hss}}}
\makeatother

\begin{document}
\title{The string/gauge theory correspondence in QCD\footnote{Lectures
  presented by KP at the 45th Winter School on Theoretical Physics,
  Feb.~2007, Schladming.}}
\author{Kasper
  Peeters\inst{1}\fnmsep\thanks{\email{k.peeters@phys.uu.nl}}
  \and Marija
  Zamaklar\inst{2}\fnmsep\thanks{\email{marija.zamaklar@durham.ac.uk}}}
\institute{Institute for Theoretical Physics, 
  Utrecht University, 
  P.O.~Box 80.195, 
  3508 TD Utrecht,\\
  The Netherlands
  \and 
  Department of Mathematical Sciences, 
  Durham University, 
  South Road, 
  Durham DH1 3LE, \\
  United Kingdom}
\abstract{Ideas about a duality between gauge fields and strings have been
around for many decades. During the last ten years, these ideas have
taken a much more concrete mathematical form. String descriptions of
the strongly coupled dynamics of semi-realistic gauge theories,
exhibiting confinement and chiral symmetry breaking, are now
available. These provide remarkably simple ways to compute properties
of the strongly coupled quark-gluon fluid phase, and also
shed new light on various phenomenological models of hadron
fragmentation. We present a review and highlight some exciting
recent developments.}
\maketitle
\markboth{Kasper Peeters and Marija Zamaklar}{The string/gauge theory correspondence in QCD}

\section{Introduction and motivation}

Although there are several well-known and clear-cut signs of strings
in gauge theories, a concrete realisation of the connection only
became available with Maldacena's conjecture of the AdS/CFT
correspondence~\cite{mald2}.  The crucial new ingredient, namely that
strings dual to gauge fields require additional dimensions, had been
anticipated by others (see e.g.~\cite{Polyakov:1997tj}), but the
discovery of D-branes in string theory~\cite{polc1} was necessary to
put this idea on a firm footing. Since 1997, the developments have
gone essentially two ways. On the one hand, a lot of effort has gone
into proving the correspondence as mathematically precise as possible,
uncovering intriguing integrable structures along the way. On the
other hand, much work has been done on trying to extend the original
conjecture from the highly supersymmetric case of~\cite{mald2} to more
realistic gauge theories, and QCD in particular.

The generalisation of the AdS/CFT correspondence to more realistic
gauge theories, also dubbed AdS/QCD or non-AdS/non-CFT, has been
important for two reasons. Firstly, it provides a rigorous basis for
many phenomenologically successful string models of strong
interactions. It explains how having a four-dimensional gauge theory
and at the same time a critical string is compatible, keeping both the
good features of phenomenological models and the good features of a
tachyon-free quantum conformally invariant string. In this
way, one obtains a new way to think about e.g.~hadron fragmentation,
and also finds new candidates to model the energy distribution
of gluon flux tubes as seen on the lattice (see e.g~\cite{Kuti:2005xg}).

Secondly, it provides a way to do dynamical strong coupling
calculations at finite temperature. The main driving force here is the
discovery of the strongly coupled quark gluon fluid at RHIC (see
e.g.~\cite{Muller:2006ee} for a review). Rather unexpectedly, the
quark-gluon fluid has provided string theorists with the long
sought-after laboratory in which strongly-coupled gauge theory
dynamics can be explored directly. In return, the AdS/CFT
correspondence provides entirely new ways to compute otherwise
hard-to-compute properties of the fluid.

By now, the literature on the AdS/CFT correspondence and its
generalisation to more realistic gauge theories is huge. Both the
attempts to develop a formal proof of the correspondence, as well as
the attempts to generalise it to realistic gauge theories, have led to
an enormous number of papers. The goal of the present review is to
give a concise overview of this vast field, focusing on the
applied part of it (the formal side of the field is developing very
quickly and there is at present unfortunately no up-to-date review;
see e.g.~\cite{Beisert:2004ry} for a somewhat older review text). Starting more or less from
scratch, we introduce the basics of the correspondence and then build
up to the exciting recent applications to the strongly-coupled
quark-gluon fluid.

\section{Old and new ideas on strings in gauge theory}
\subsection{Signs of strings}

\begin{figure}[ht]
\begin{center}
\includegraphics[width=.6\textwidth]{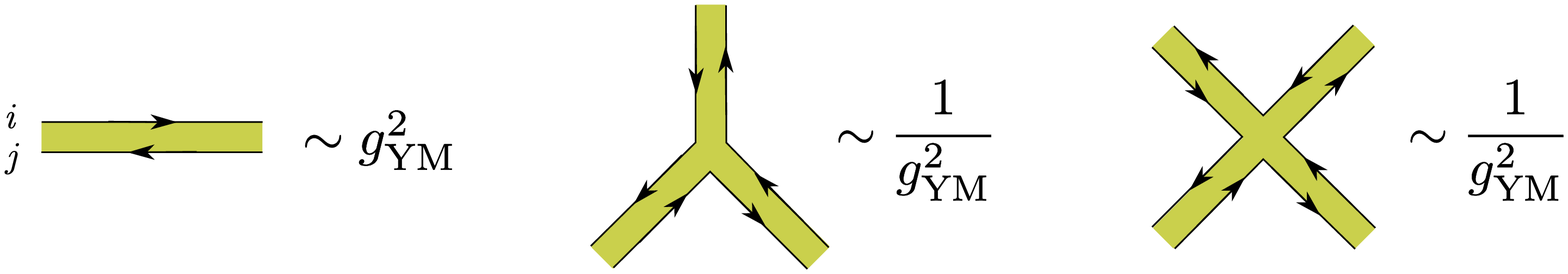}\hspace{5em}
\raisebox{1ex}{\includegraphics[width=.15\textwidth]{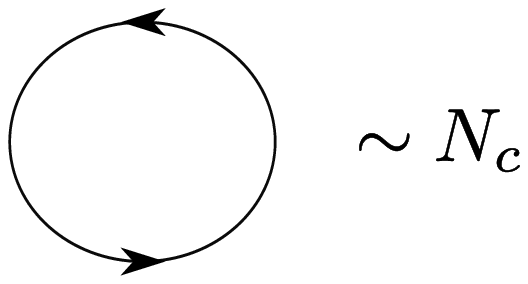}}
\end{center}
\caption{Feynman rules for SU($N$) gauge theory in the double-line
  notation. The powers of~$g_{\text{YM}}$ come from the propagator and
  vertices, while every closed loop of a black edge yields a
  power~$N_c$.\label{f:double-line}}
\end{figure}

Long before string theory developed into a theory of quantum gravity,
it was known that there are ``stringy'' features present in gauge
theories. One of the mathematically most explicit ways in which
strings can be seen to appear is in the limit in which the number of
colours~$N_c$ is taken large. In this limit, the perturbation series
organises itself in an intriguing way~\cite{'tHooft:1973jz}. Let us
discuss for simplicity the pure-gauge theory, given by the action
\begin{equation}
S = \frac{1}{4\,g_{\text{YM}}^2} \int\!{\rm d}^4x\, \Tr \big( F_{\mu\nu} F^{\mu\nu}
\big)\,,\quad
F_{\mu\nu} = \partial_\mu A_\nu - \partial_\nu A_\mu + [A_\mu,A_\nu]\,,\quad
(A_{\mu})_{ij} = A_{\mu}^a\; (T^a)_{ij}\,.
\end{equation}
Here~$(T^a)_{ij}$ are the generators of $\text{SU}(N_c)$. It is
convenient to use a graphical notation in which the two indices~$i,j$
are each given their own line, the so-called ``double-line
notation''. The Feynman rules are then as in figure~\ref{f:double-line}.

\begin{figure}[t]
\begin{center}
\includegraphics[height=.4\textwidth]{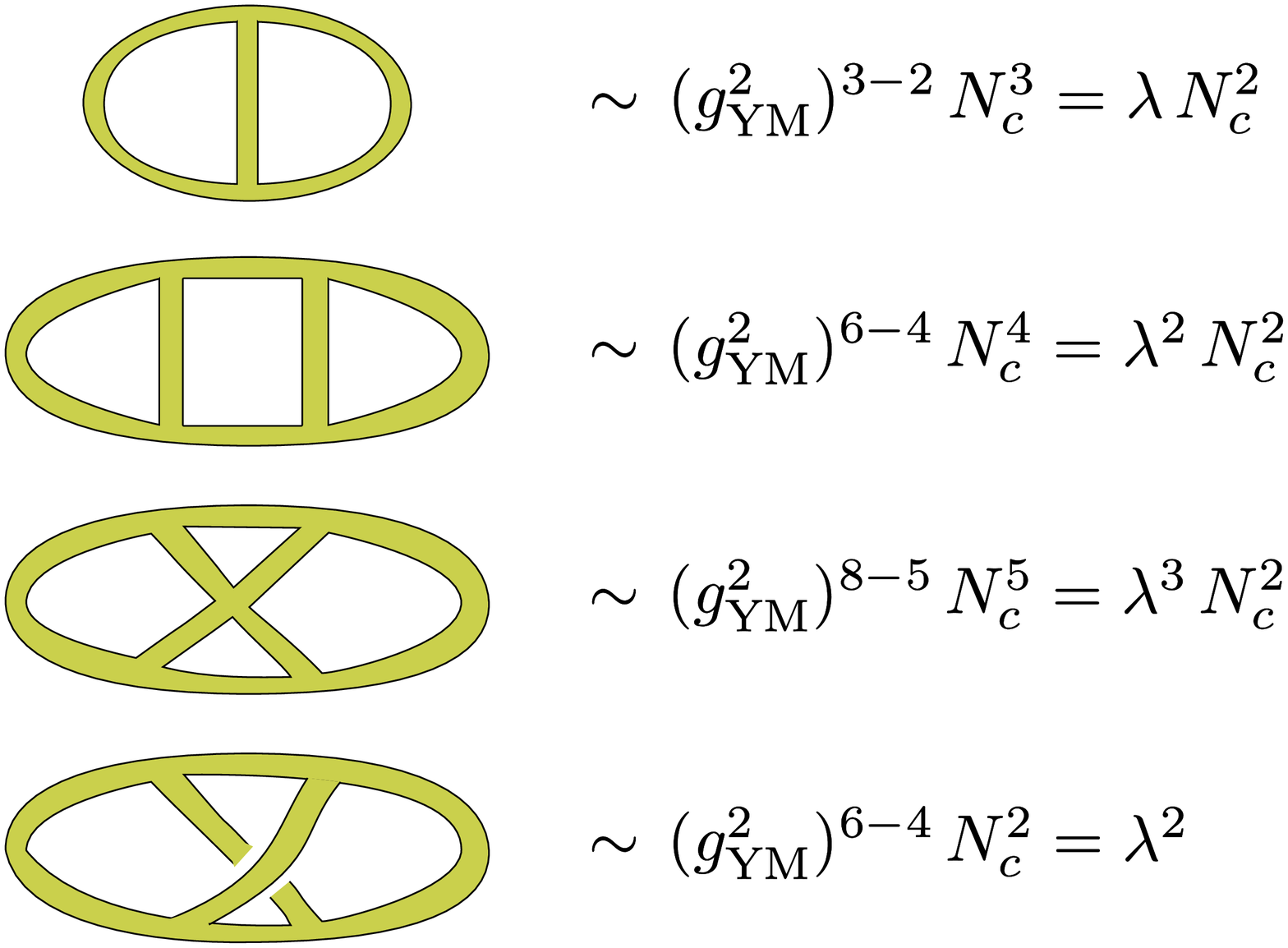}
\hspace{4em}\raisebox{.5ex}{\includegraphics[height=.38\textwidth]{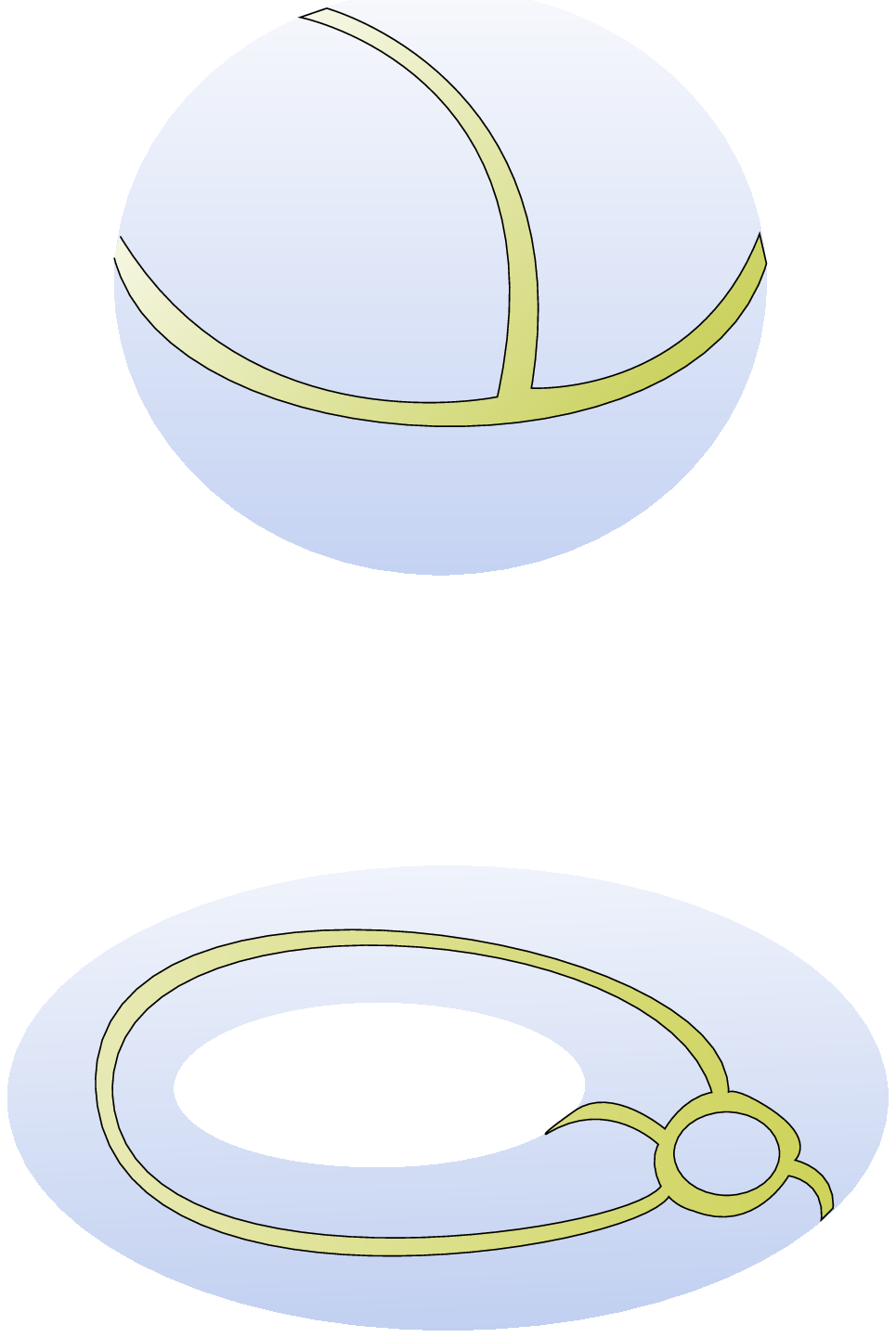}}
\end{center}
\vspace{1ex}
\caption{Counting powers of~$g_{\text{YM}}$ and $N_c$. By combining
  the coupling constant and the number of colours into the so-called
  't~Hooft coupling~$\lambda = g_{\text{YM}}^2 N_c$, one sees
  that~$N_c^{-2}$ counts the degree of non-planarity (left). Planar diagrams
  can be drawn planar on a sphere, while non-planar diagrams require
  higher-genus surfaces in order to be drawn as planar graphs (right).\label{f:genusdiagrams}}
\end{figure}

Let us look at some simple diagrams and count the powers
of~$g_{\text{YM}}$ and $N_c$. A number of examples has been drawn in
figure~\ref{f:genusdiagrams}. They serve to illustrate the following
result: by combining the coupling constant and the number of colours
into the combination~$g_{\text{YM}}^2 N_c$, one can show that the
degree of non-planarity of a graph is counted by~$N_c^{-2}$. The planar
diagrams all scale like~$N_c^2$, while non-planar diagrams are
suppressed by at least one factor of~$N_c^{-2}$. The
combination~$g_{\text{YM}}^2 N_c$ is called the ``'t~Hooft
coupling'' and commonly denoted with~$\lambda$. By keeping this
coupling fixed, we see that the large-$N_c$ expansion is an expansion
in the degree of non-planarity, with planar diagrams being the most
dominant.

The connection to strings now comes about by trying to figure out how
to draw non-planar diagrams as planar ones, making use of  surfaces of
non-trivial topology. The last diagram of figure~\ref{f:genusdiagrams}
is non-planar, but it can be drawn as a planar diagram \emph{on the
torus}. This is a general phenomenon: diagrams with a higher degree of
non-planarity can be drawn as planar diagrams on surfaces of higher
\emph{genus}. In this way, we see that there is a kind of dual
relationship between Feynman diagrams of~$\text{SU}(N_c)$ gauge theory
and two-dimensional surfaces. 

These surfaces suggest that there is something stringy about the
Feynman diagram expansion. The diagrams can be seen as the duals of
Euclidean versions of closed-string world-sheets, with one direction
labelling the direction along the string and one direction labelling
the evolution in time. Unfortunately, it is very hard to make this
relation more precise, or in fact to use it to compute Feynman
diagrams (a recent fresh approach to this program can be found
in~\cite{Gopakumar:2003ns}). One of the main puzzles is that, even
though we have identified the meaning of~$N_c$, we still do not know
what is the meaning of the 't Hooft coupling~$\lambda$ in terms of the
string theory.

\begin{figure}[t]
\vspace{4ex}
\begin{center}
\psfrag{Msq}{$M^2$}
\psfrag{J}{$J$}
\psfrag{a2}{\smaller\smaller\smaller\hspace{-6ex} $a_2$}
\psfrag{K1430}{\smaller\smaller\smaller $K(1430)$}
\psfrag{rho770}{\smaller\smaller\smaller\hspace{-8ex} $\rho(770)$}
\psfrag{K892}{\smaller\smaller\smaller\hspace{3ex} $K(892)$}
\psfrag{rho31690}{\smaller\smaller\smaller\hspace{-8ex} $\rho_3(1690)$}
\psfrag{K1780}{\smaller\smaller\smaller\hspace{3ex} $K(1780)$}
\psfrag{a42040}{\smaller\smaller\smaller\hspace{-8ex} $a_4(2020)$}
\psfrag{K2045}{\smaller\smaller\smaller\hspace{3ex} $K(2045)$}
\psfrag{rho52350}{\smaller\smaller\smaller\hspace{-8ex} $\rho_5(2350)$}
\psfrag{K2380}{\smaller\smaller\smaller\hspace{2ex} $K(2380)$}
\includegraphics[height=4cm]{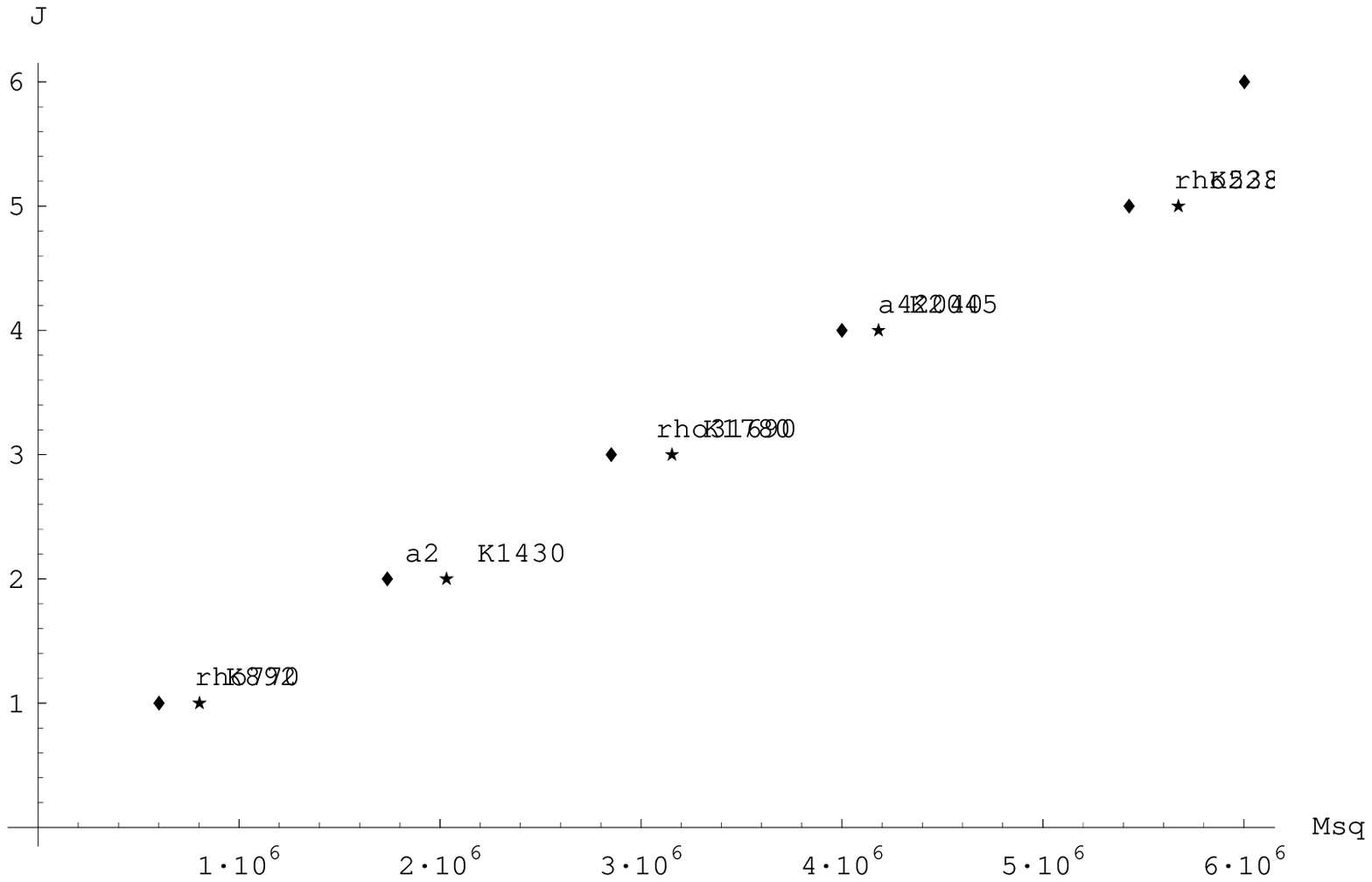}\qquad\qquad
\psfrag{Msq}{$M^2$}
\psfrag{n}{$n$}
\psfrag{rho770}{\smaller\smaller\smaller\hspace{-8ex} $\rho(770)$}
\psfrag{rho1450}{\smaller\smaller\smaller\hspace{-8ex} $\rho(1450)$}
\psfrag{rho1700}{\smaller\smaller\smaller\hspace{-8ex} $\rho(1700)$}
\psfrag{rho1900}{\smaller\smaller\smaller\hspace{-8ex} $\rho(1900)$}
\includegraphics[height=4cm]{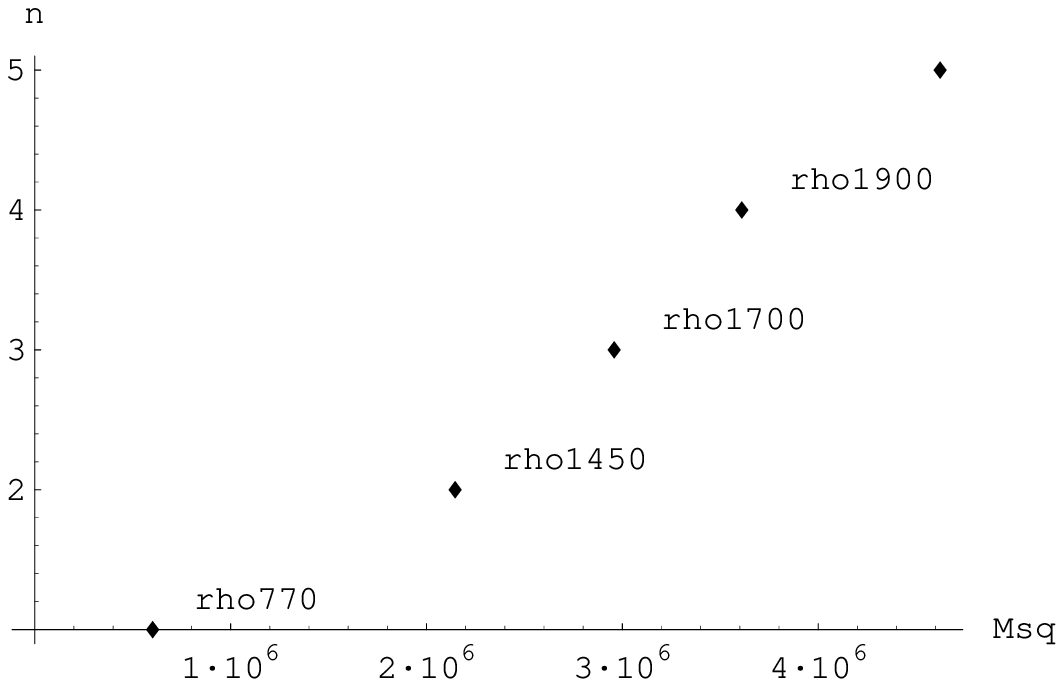}
\end{center}
\caption{Regge trajectories of mesons relate the spin~$J$ to
  the mass-squared~$M^2$ (left) or the excitation level~$n$ to the
  mass-squared (right) \cite{PDBook}.\label{f:Regge}}
\end{figure}

String features of gauge theory are, however, not only visible in
perturbation theory around weak coupling. They are also manifest in
the \emph{strong coupling} regime, where there are obvious signs of
the presence of string-like forces. This is in particular clear by
looking at the spectrum of mesons or baryons. These spectra display,
to good approximation, a linear relation between the spin~$J$ and the
mass-squared~$M^2$,
\begin{equation}
J = \alpha + \alpha' \, M^2 \,,
\end{equation}
with different trajectories distinguished by different
intercepts~$\alpha$, yet having an almost identical slope~$\alpha'$.
A similar linear Regge relation can be seen by keeping the spin fixed
but looking instead at the excitation level~$n$ (see
figure~\ref{f:Regge}). These linear relations are easily obtained from
a string-like model of mesons, in which mesons are seen as massive
quarks or di-quarks connected by a relativistic string (see
e.g.~\cite{Wilczek:2004im} for a recent review). While suggestive,
this again leaves us with the problem of understanding the meaning of
the 't Hooft coupling.

\subsection{The AdS/CFT correspondence}

The AdS/CFT correspondence, first conjectured by
Maldacena~\cite{mald2} based on earlier work by
Polyakov~\cite{Polyakov:1997tj}, gives a concrete handle on the
meaning of the 't Hooft coupling~$\lambda$, as well as a concrete
answer to the question ``where is the QCD string?''. The key ingredient is
that the strings of the theory dual to gauge theory do not reside in
our four-dimensional world, but are instead contained in an auxiliary,
higher-dimensional curved space-time. There exists a specific and
well-defined map between the calculations which one can do with the
strings in this higher dimensional space and the calculations one can
do in the four-dimensional gauge theory; we will return to this map
shortly. 

\begin{figure}[t]
\begin{center}
\hbox{\vbox{\hbox{\includegraphics[width=.4\textwidth]{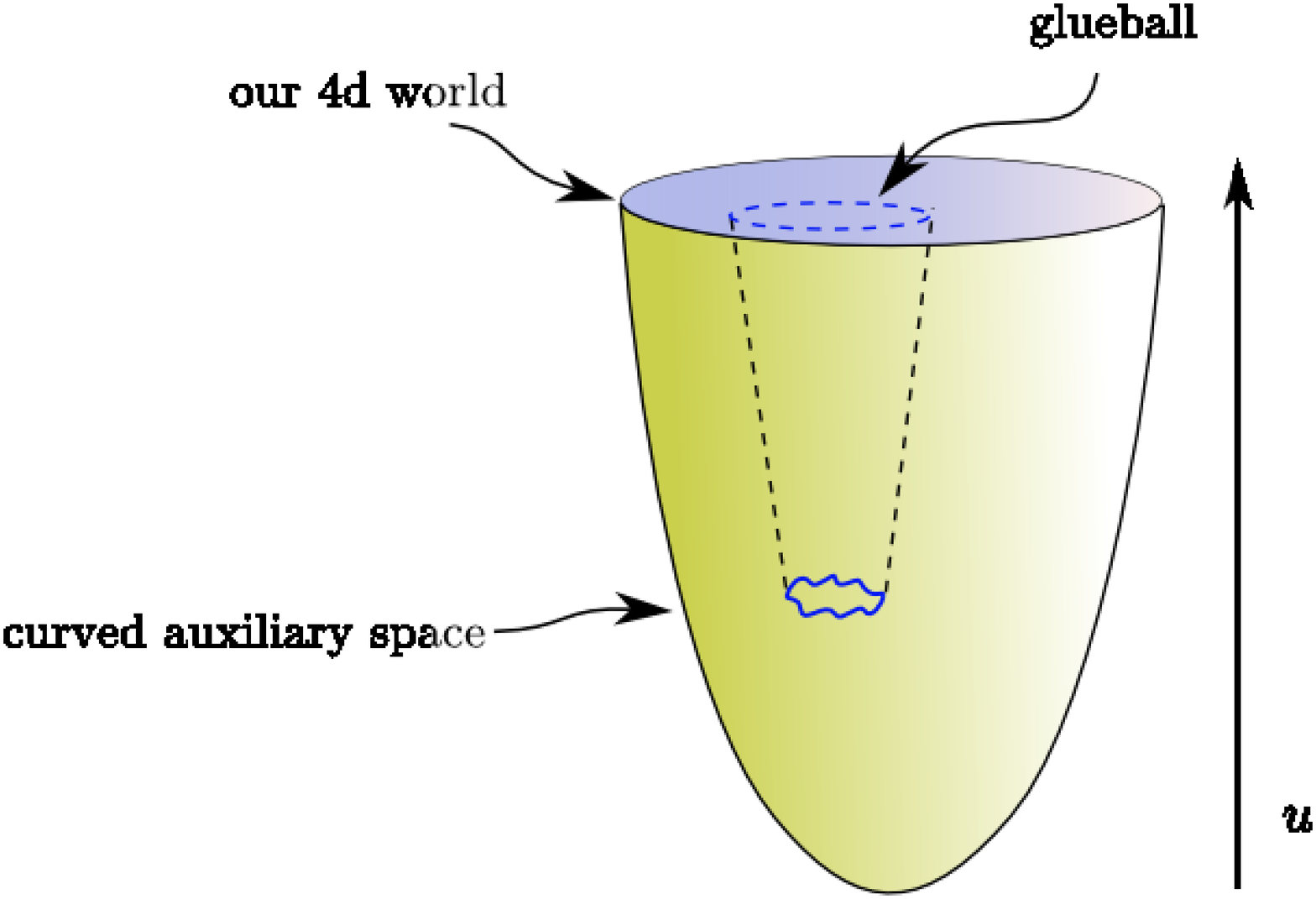}}\hbox{\hspace{10em}conformal}\hbox{\hspace{10em}\smaller
    anti de-Sitter}}\hspace{4em}
\vbox{\hbox{\includegraphics[width=.2\textwidth]{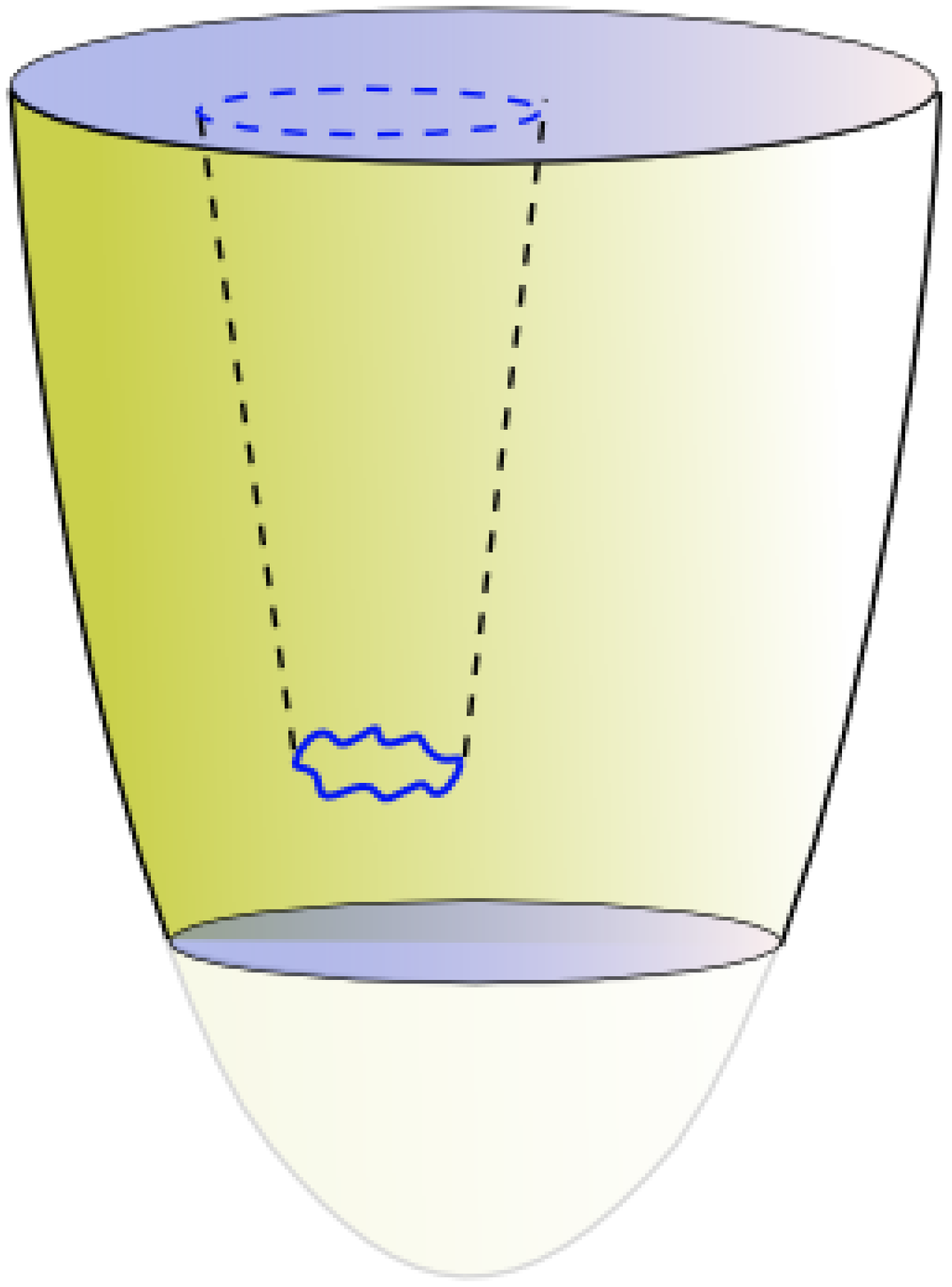}}\hbox{~~~~~~~~confining}\hbox{\smaller
      ~~~~~extra scale (``wall'')}}\hspace{1em}
\vbox{\hbox{\includegraphics[width=.2\textwidth]{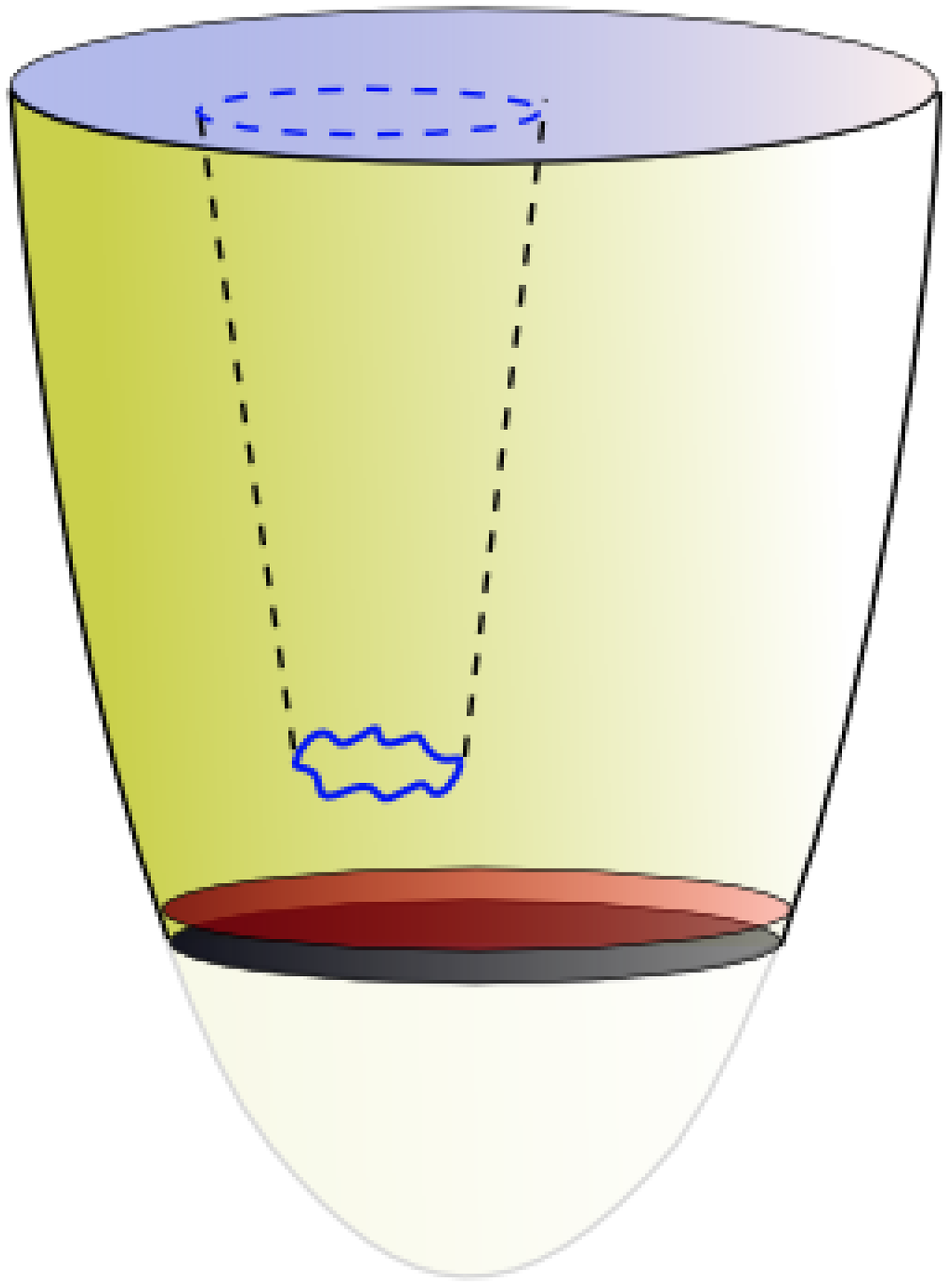}}\hbox{~~~~~~~~~thermal}\hbox{~~\smaller Hawking-radiating
    black hole}}}
\end{center}
\caption{Symbolic depiction of the basic idea of the AdS/CFT
  correspondence: the string theory dual to gauge theory is
  higher-dimensional. The string lives in a curved space-time, and
  there is a specific map which relates the physics of the string to
  the physics in our four-dimensional world. More recent extensions of
  this conjecture have produced string geometries dual to confining
  and thermal theories (right).\label{f:symbolic}}
\end{figure}

In the original and best-understood form of this conjecture, the
curved string space-time is five-dimensional Anti-de-Sitter space
times a five-sphere, $\text{AdS}_5\times S^5$, and the dual gauge
theory lives on its boundary, ${\mathbb R}\times S^3$. Importantly, it
gives a precise relation between the 't~Hooft coupling and parameters
on the string theory side,
\begin{equation}
\label{e:lambdastring}
\lambda = \frac{R^4}{(l_{\text{string}})^4}\,,
\end{equation}
where~$R$ is the radius of curvature of the $\text{AdS}_5$ space-time,
and $l_{\text{string}}$ is the string length (we will also
use~$\alpha' = (l_{\text{string}})^2$). When the curvature radius is
very large compared to the typical size of the string, $R\gg
l_{\text{string}}$, only the massless states in the spectrum of the string
survive, and one ends up with a \emph{gauge/gravity} correspondence.
In this regime, where string theory is best under control, the gauge
theory is strongly coupled. By now, it is also known which qualitative
features of the string geometry have to be modified in order to
generalise the correspondence to confining and thermal gauge theories
(see figure~\ref{f:symbolic}). In all of these cases, we have a
relation similar to~\eqref{e:lambdastring}, stating that weakly
coupled strings are dual to a strongly coupled gauge theory.

\begin{figure}[t]
\begin{center}
\includegraphics[width=.8\textwidth]{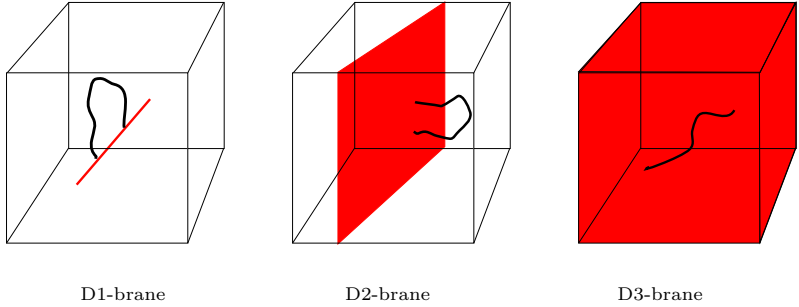}\\
{{\smaller D1-brane}~~~~~~~~~~~~~~~~~~~~~~{\smaller
	 D2-brane}~~~~~~~~~~~~~~~~~~~~~~~{\smaller D3-brane}}
\end{center}
\caption{The theory of open strings contains various heavy objects in
  its spectrum, which can be interpreted as boundary conditions on the
  endpoints of the light open strings.\label{f:dbranes}}
\end{figure}

How does one arrive at such an idea? It arose by carefully analysing
the dynamics of D-branes in string theory. D-branes are very massive
objects in the spectrum of relativistic strings~\cite{polc1}. To first
approximation, one can think of them as fixed hypersurfaces in
space-time, which yield Dirichlet boundary conditions for light
strings (see figure~\ref{f:dbranes}). Depending on which string theory
one is working with, they come in various dimensionalities. In the
type-IIB string theory, one finds in particular 3+1~dimensional D3-branes. The key
idea is now to look at a stack of~$N_c$ of these. Each of the~$N_c$
D3-branes couples to the gravitational degrees of freedom with a
strength~$g_s$, so in total the strength of the gravitational
distortion depends on~$g_s N_c$. There are now two regimes in which we
can consider the system.

When $g_s N_c \ll 1$, space-time is nearly flat. The fluctuations of
the D-brane are described by open strings. There are also closed
strings, but they decouple from the open strings because the string
coupling is weak. If we focus on low-energy excitations, these closed
string degrees of freedom describe a decoupled linearised gravity
theory. The open string low-energy degrees of freedom are described by
the open string effective action, which is a super-Yang-Mills theory
on the D-brane.

When $g_s N_c \gg 1$, back-reaction of the branes on the background
becomes important, and space-time is curved. This curvature is
described in terms of the effective action for closed strings,
i.e.~the supergravity theory in the full space-time (the ``bulk''). If
we look at low-energy excitations as seen from an observer at
infinity, these are essentially the excitations close to the horizon,
as those are strongly red-shifted. These excitations cannot easily
escape to the asymptotic region. Moreover, low-energy excitations from
the asymptotic regime find it hard to probe the near-horizon regime,
because their wavelength is so large. Therefore, we again find two
decoupled theories, one describing linearised gravity in the
asymptotic region, and one describing gravity in the near-horizon
region.

Thus, the asymptotic region is, both in the $g_s N_c\ll 1$ and in the
$g_s N_c\gg 1$ regime, described by decoupled linearised gravity.
Since it is decoupled, we are left with the effective action for
\emph{open} strings in flat space-time when $g_s N_c\ll 1$, and the
effective action for \emph{closed} strings in the near-horizon region
when $g_s N_c \gg 1$. The statement of the string/gauge theory
correspondence is now that the gauge theory describing the open
strings on the D3-brane and the supergravity theory describing the
closed strings in the near-horizon regime of the curved D3-brane
supergravity solution are two \emph{different} descriptions of the \emph{same}
physical system.

More precisely, the open string degrees of freedom are described by a
four-dimensional Dirac-Born-Infeld action for the gauge fields (see
e.g.~\cite{Fradkin:1985qd}), which at lowest order in derivatives
reduces to $N=4$ super-Yang-Mills,
\begin{equation}
S = \frac{1}{g_{\text{YM}}^2}\int\!{\rm d}^4 x\, \Tr\Big( \frac{1}{4} F_{\mu\nu}
F^{\mu\nu} + \frac{1}{2}(D_\mu\phi_{ij})^2
+ \frac{1}{2}\bar{\chi}_i\slashed{D}\chi_i  
- \frac{1}{2}\bar{\chi}_{i}[\phi_{ij},\chi_{j}]
- \frac{1}{4}[\phi_{ij}, \phi_{kl}][\phi_{ij}, \phi_{kl}]\Big)
\end{equation}
with 6 adjoint scalars $\phi^{(ij)}$, a gauge field
$A_\mu$ and 4 chiral adjoint fermions $\chi_i$. The closed strings
are, for large values of~$\lambda$, described by supergravity
excitations around the near-horizon geometry of the extremal
D3-brane. The metric of the D3-brane is~\cite{horo4},
\begin{equation}
\label{e:N4YM}
{\rm d}s^2 = \left(1 + \frac{R^4}{u^4}\right)^{-1/2} \left(
- {\rm d}t^2 + {\rm d}x_1^2 + {\rm d}x_2^2 + {\rm d}x_3^2 \right)
+ \left(1 + \frac{R^4}{u^4}\right)^{1/2} 
  \left( {\rm d}u^2 + u^2\, {\rm d}\Omega_5^2\right)\,.
\end{equation}
This has a horizon at~$u=0$ (these are isotropic coordinates),
as~$g_{tt}\rightarrow 0$ there. There is also a five-form field, whose
charge is fixed by the supergravity equations of motion to
\begin{equation}
\int_{S^5}\, F_5 = N_c\,,\qquad\text{and}\qquad 4\pi g_s\, N_c\, l_s^4
= R^4\,,
\end{equation}
where~$g_s$ is the string coupling constant.  The conjecture says that
we should go close to the horizon of the D3-brane, where we end up
with the metric
\begin{equation}
\label{e:AdS5S5}
{\rm d}s^2 = \frac{u^2}{R^2} \left(
- {\rm d}t^2 + {\rm d}x_1^2 + {\rm d}x_2^2 + {\rm d}x_3^2 \right)
+ \frac{R^2}{u^2}\,{\rm d}u^2 ~~+
~~\phantom{\Big|}R^2\, {\rm d}\Omega_5^2\,.
\end{equation}
This is the~$\text{AdS}_5\times S^5$ metric on which the string theory
dual to the gauge theory lives (one often also uses a coordinate $z =
1/u$ which makes the conformal equivalence of $\text{AdS}_5$ to flat
space come out more clearly). More formally, the correspondence thus
states that type-IIB string theory on the background~\eqref{e:AdS5S5}
is equivalent to the gauge theory described by~\eqref{e:N4YM} on the
$\mathbb{R}\times S^3$ boundary.\footnote{Strictly speaking, the
  boundary of~\eqref{e:AdS5S5} is~$\mathbb{R}\times
  \mathbb{R}^3$. Both $\mathbb{R}\times S^3$ and~$\mathbb{R}\times \mathbb{R}^3$ can be viewed as the
  boundary of~$\text{AdS}_5$ as they are related by a conformal
  transformation if the point at infinity in~$\mathbb{R}^3$ is included.}

\subsection{The holographic dictionary}
\label{s:holodict}

Let us now discuss in some detail how computations done on the
string theory side can be related to those which one would like to do
on the gauge theory side. The relation between correlation functions
on the two sides which we will discuss below was first proposed
in~\cite{Gubser:1998bc,Witten:1998qj}. The connection is made at the
level of the generating functional for correlation functions. In a
nutshell, the connection is written as
\begin{multline}
Z_{\text{string}} = \exp\bigg[ - S_{\text{sugra}}\Big({\phi}(t,\vec{x};\, u=\infty) =
u^{\Delta-4}{\phi_0(t,\vec{x})}\Big)\bigg] + \ldots \\[1ex]
= \left\langle T\,\exp {\displaystyle \int\!{\rm d}^4x\,
  {\phi_0(t,\vec{x})}\,{\cal O}(t,\vec{x})}\right\rangle_{\text{field theory}}\,.
\end{multline}
Let us explain this in some detail. The field~$\phi(t,\vec{x}; u)$ is
a field on the~$\text{AdS}_5\times S^5$ geometry, for instance a
graviton fluctuation. It is a function of all five coordinates, in
particular the radial direction~$u$ of the metric~\eqref{e:AdS5S5}.
Near the boundary, the fluctuations scale with~$u$ according to their
conformal dimension~$\Delta$. The first line of the expression above
instructs us to evaluate the string partition function, which to first
order is given by the exponent of the supergravity action, given a
certain boundary behaviour of the field~$\phi$. The correspondence now
states that the result is the same as computing the field theory
expectation value of the exponent of a specific operator~${\cal O}$,
coupled to $\phi_0(t,\vec{x})$, now interpreted as a source.

Correlation functions on the gauge theory side now follow simply by
computing repeated derivatives with respect to the source,
\begin{equation}
\label{e:corrcorr}
\frac{\delta^n\, Z_{\text{string}}}{\delta {\phi_0(t_1,\vec{x}_1)} \cdots
\delta {\phi_0(t_n, \vec{x}_n)}} 
 = \Big\langle T\, {\cal O}(t_1,\vec{x}_1)\cdots {\cal O}(t_n,
 \vec{x}_n)\Big\rangle_{\text{field theory}}\,.
\end{equation}
It is illustrative to discuss this method of computing correlators for
at least one simple example. So let us discuss the canonical one,
namely the two-point function of a massive scalar field (see
e.g.~\cite{Freedman:1998tz,Aharony:1999ti,Son:2002sd}). The action for a massive
scalar field on $\text{AdS}_5$ is given by (using~$z=1/u$)
\begin{equation}
S = \frac{1}{g_s^2}\int_0^{\infty}\!{\rm d}z{\rm d}^4x\, \frac{1}{z^3} \Big[ 
(\partial_z\phi)^2 + (\partial_{\mu}\phi)^2 + \frac{m^2 R^2}{z^2}\phi^2\Big]\,.
\end{equation}
If we want correlation functions in the boundary theory in momentum
space, we should make an expansion of the field~$\phi(x^\mu; z)$ in
terms of $z$-dependent modes~$f_{k}(z)$ and boundary fields~$\phi_0(k_\mu)$,
\begin{equation}
\phi(x^\mu; z) = \int\!\frac{{\rm d}^4k}{(2\pi)^4}\, e^{ik\cdot x}
f_k(z)\, \phi_0(k_\mu)\,.
\end{equation}
This yields the action
\begin{equation}
S = \frac{1}{g_s^2}\int_0^{\infty}\!{\rm d}z \int\!\frac{{\rm d}^4k
  {\rm d}^4k'}{(2\pi)^8}
\frac{1}{z^3}\Big[ \partial_z f_k \partial_z f_{k'} + k^2 f_k f_{k'} + \frac{m^2
	 R^2}{z^2} f_k f_{k'}\Big] \phi_0(k_\mu) \phi_0(k'_\mu) \delta^{(4)}(k+k')\,.
\end{equation}
By partially integrating the first term in the action with respect to~$z$, we
find the equation of motion for~$f_k$ plus a boundary term,
\begin{equation}
\label{e:Sonshell}
S_{\text{on shell}} = \frac{1}{g_s^2}\int\!\frac{{\rm d}^4k{\rm d}^4k'}{(2\pi)^8}\,
\delta^{(4)}(k+k')\phi_0(k_\mu)\phi_0(k'_\mu)\,\frac{f_{k'}\partial_z f_k}{z^3}\Big|^{\infty}_{\epsilon}\,,
\end{equation}
where we have regulated the boundary using a small
parameter~$\epsilon$. By taking two variational derivatives with
respect to the ``source'' $\phi_0(k)$, as in~\eqref{e:corrcorr}, we thus
immediately read off an expression for the two-point function in $N=4$
super Yang-Mills.

In order to compute it, we need to solve the equation of motion for
the $z$-dependent part of the solution. It is given by
\begin{equation}
\label{e:bulkEOM}
f''_k - \frac{3}{z} f_k' - \left( k^2 + \frac{m^2 R^2}{z^2}\right) f_k
= 0\,.
\end{equation}
This equation has the general solution
\begin{equation}
\label{e:fsol}
f_k = C_1\, z^2\, I_{\Delta-2}(kz) + C_2\, z^2\,
K_{\Delta-2}(kz)\,,\qquad \Delta=2 + \sqrt{4+m^2R^2}\,,
\end{equation}
where~$I_{\Delta-2}$ and $K_{\Delta-2}$ are modified Bessel functions
of the first and second kind.  Imposing regularity in the interior
at~$z=\infty$ forces us to set~$C_1=0$; normalising the remaining
solution to one at~$z=\epsilon$ (i.e.~near the boundary), we end up
with
\begin{equation}
f_k(z) = \frac{z^2\, K_{\Delta-2}(kz)}{\epsilon^2\, K_{\Delta-2}(k\epsilon)}\,.
\end{equation}
Inserting this in~\eqref{e:Sonshell}, we find a contribution from
the~$z=\epsilon$ boundary which is given by
\begin{equation}
\langle {\cal O}(k) {\cal O}(k') \rangle
 = \epsilon^{2(\Delta - 2)} \delta^{(4)}(k+k')k^{2(\Delta-2)} 2^{1-2(\Delta-2)}
 \frac{\Gamma(3-\Delta)}{\Gamma(\Delta-1)}(\Delta-2) + \ldots\,,
\end{equation}
plus contact terms.  The details are not very important at this stage
(and there are several subtleties in identifying which are the gauge
theory operators~${\cal O}$ and how to normalise them properly, which we
have not discussed). What is important is the general logic. A priori,
there are two independent solutions for~$f_k$ near the
boundary. Imposing regularity of the supergravity solutions in the
interior removes one of the coefficients and leaves only an overall
normalisation. The boundary term in the action then leads to the
correlator of the dual gauge theory operators. This logic holds also
for more complicated string/gauge theory duals. 

The spectrum of the bulk modes satisfying equations such
as~\eqref{e:bulkEOM} is in one-to-one relation with the spectrum of
gauge invariant operators in the boundary theory. This connection is
at the basis of computations of gauge theory spectra (for instance of
glueballs~\cite{Csaki:1998qr,deMelloKoch:1998qs}) by solving
supergravity spectral problems. We will see this connection for mesons
in more detail in later sections.

Apart from computing correlators or the spectrum of physical states,
one is often also interested in vacuum expectation values of certain
operators. The value of a bulk field at the boundary is related, by
the holographic dictionary, to a chemical potential. By varying this
leading component~$\mu$ of the boundary field~$\phi_0(t,\vec{x};z)$
(the ``non-normalisable mode), we can compute the expectation value of
the dual operator in a particular state. The logic is similar to the
one above for correlators. By varying the boundary field, one ends up
with an equation of motion plus a boundary term,
\begin{equation}
\delta S = \text{EOM}_{\text{bulk}} + \bigg(\frac{\partial
 S}{\partial \phi} \frac{\delta \phi}{\delta \mu} \delta \mu \bigg)_{\text{boundary}} 
 = \text{EOM}_{\text{bulk}} + \delta \mu \langle O_{\mu} \rangle \, .
\end{equation}
If one writes this in a series expansion near the boundary, one finds
that whereas $\mu$ is the coefficient of the leading term, the
expectation value~$\langle {\cal O}\rangle$ is related to the
subleading or ``normalisable'' coefficient of the bulk field.

One final comment concerns the meaning of the radial direction. The
cut-off~$\epsilon$ is an ultra-violet cut-off in the dual gauge
theory. The radial direction~$u$ is therefore related to the energy
scale in the dual gauge theory. Strongly coupled infrared physics in
the gauge theory is related to string phenomena which happen deep
inside the bulk of the supergravity solution.

\section{String duals with confinement and chiral symmetry breaking}
\subsection{From Coulomb potentials to confining potentials}
\label{s:coulomb_to_confining}

Having access to a string dual of $N=4$ super-Yang-Mills, the question
of course arises as to whether it is possible to use the same idea to
construct string duals of more realistic gauge theories. In
particular, it would be interesting to find a string dual to a gauge
theory which exhibits confinement. 

\begin{figure}[t]
\hspace{2em}
\includegraphics[width=.25\textwidth]{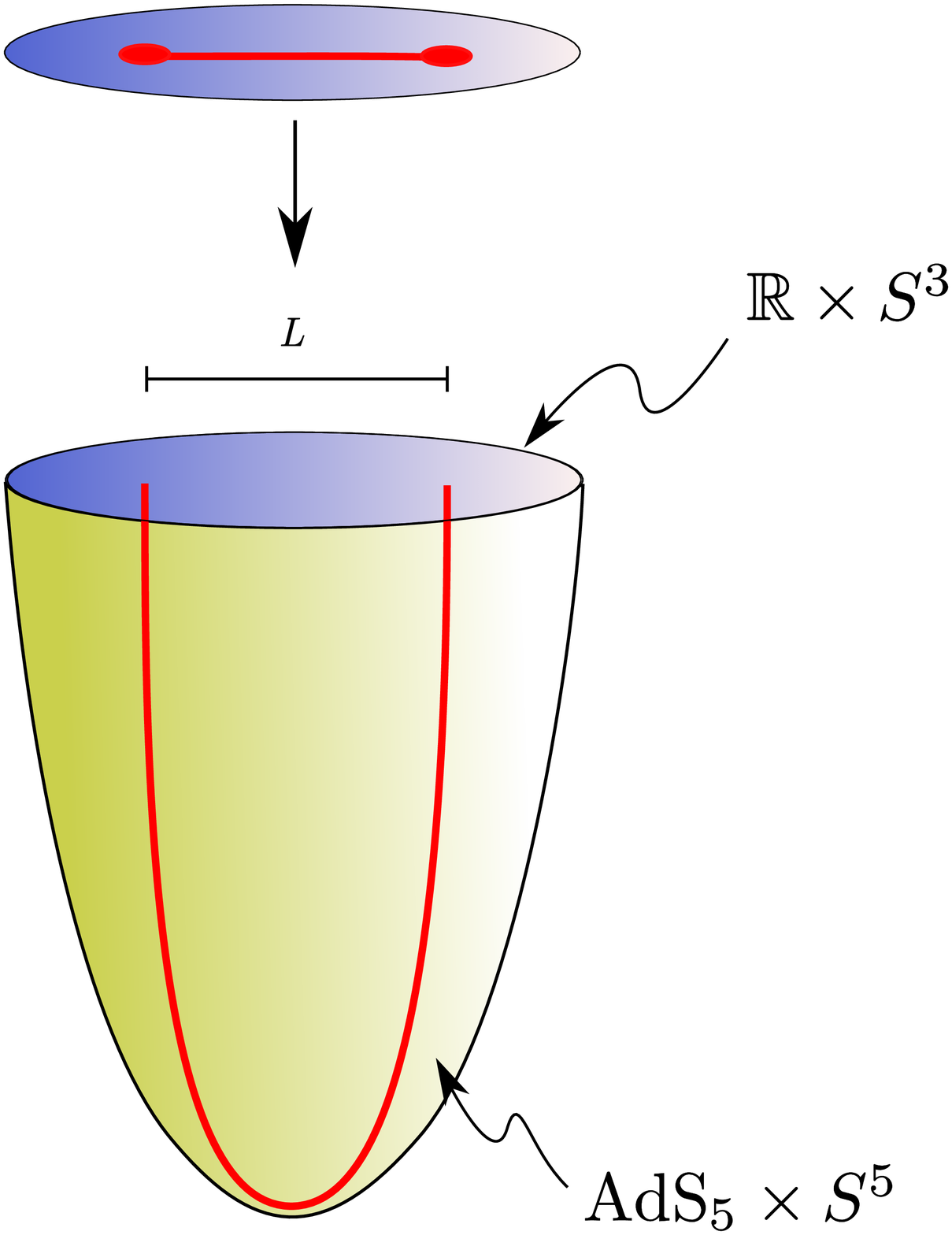}
\hspace{7em}\raisebox{4ex}{\includegraphics[width=.5\textwidth]{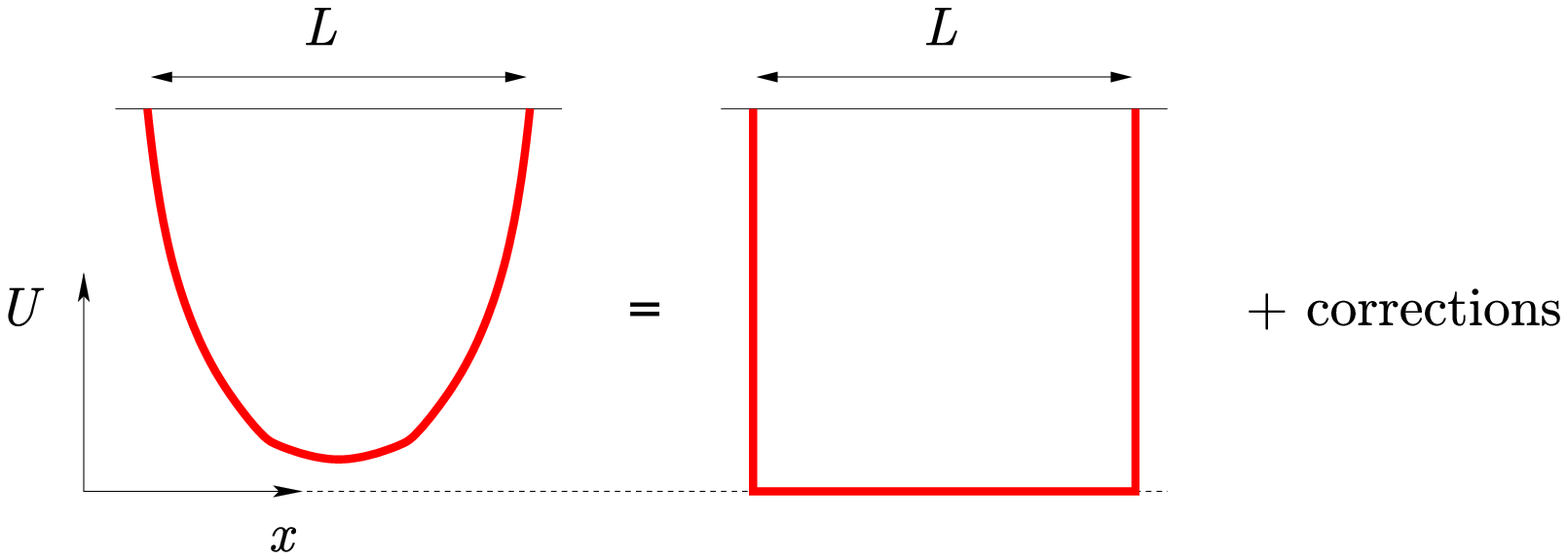}}
\caption{In order to get a confining quark/anti-quark potential, the
  Wilson loop in the gauge theory should exhibit an area law. This
  requires that the string stretching in the higher-dimensional space
  has a long horizontal segment. Conformal invariance prevents this
  (left), but there are other metrics which do allow for such a
  configuration (right).\label{f:wloop}}
\end{figure}

Recall that a signal for confinement can be found in the behaviour of
Wilson loop. For~$N=4$ super-Yang-Mills, the string configuration dual
to a Wilson loop is a string world-sheet ending with its boundary on
the Wilson loop; a slice at constant time is drawn in
figure~\ref{f:wloop}. The energy of this string configuration can be
computed, and turns out to scale as~\cite{Maldacena:1998im}
\begin{equation}
E \sim L^{-1}\,,
\end{equation}
i.e.~it exhibits a Coulomb potential. In order to obtain a string dual
to a confining theory, we somehow need to construct a curved string
background which is such that the U-shaped string has a long
horizontal segment with an effective non-zero string tension. For
the~$\text{AdS}_5\times S^5$ geometry, this long horizontal segment is
missing, and we therefore do not obtain the linear scaling of energy
with distance.

Following~\cite{Kinar:1998vq}, we can investigate the generic
behaviour of~$E$ as a function of the endpoint separation~$L$ as
follows. Consider a generic background metric, with a time-like
direction~$t$, space-like directions~$x^i$ and a radial
direction~$u$, together with a number of transverse coordinates~$x_T$
which do not play a role. The line element is given by
\begin{equation}
{\rm d}s^2 = -g_{tt}(u)\,{\rm d}t^2 + g_{xx}(u)\big( {\rm d}x^2 +
{\rm d}y^2 + {\rm d}z^2\big)
+ g_{uu}\,{\rm d}u^2 + g_{TT}(u)\,{\rm d}x_T^2\,.
\end{equation}
The string will be a U-shape extending in the energy direction~$u$ and
the space direction~$x$, as in figure~\ref{f:wloop}. In the
gauge~$\tau=t$, $\sigma=x$, the action for this string is
\begin{equation}
S_{\text{Wilson}} = \int\!{\rm d}\tau{\rm d}\sigma\,{\cal L}
= T\,\int\!{\rm d}x \sqrt{ g_{tt}(u) g_{xx}(u) +
  g_{tt}(u) g_{uu}(u)
  \left(\frac{\partial u}{\partial x}\right)^2 }\,.
\end{equation}
Making use of the fact that this expression does not depend explicitly
on~$x$, one can compute the conserved ``Hamiltonian''  associated to
translations in~$x$, and from there compute
\begin{equation}
\frac{\partial u}{\partial x} = \pm 
 \sqrt{\frac{g_{tt}(u)}{g_{uu}(u)}} 
\sqrt{\frac{g_{tt}(u) g_{xx}(u) - g_{tt}(u_0)
	 g_{xx}(u_0)}{g_{tt}(u_0) g_{xx}(u_0)}}\quad
\rightarrow
\quad
{\cal L}_{\text{on shell}} = \frac{g_{tt}(u) g_{xx}(u)}{\sqrt{g_{tt}(u_0) g_{xx}(u_0)}}\,.
\end{equation}
Here~$u_0$ is the lowest point reached by the string, where it is
assumed that~$g_{tt}g_{xx}$ has a minimum or $g_{tt} g_{uu}$ diverges.
This is all that is needed in order to compute the four-dimensional
length~$L$ and the energy~$E$,
\begin{equation}
L = \int\!{\rm d}x = \int\!{\rm d}u\, \frac{\partial x}{\partial u}\,,\qquad
E = \int\!{\rm d}x\,{\cal L} = \int\!{\rm d}u\,\frac{\partial x}{\partial u}{\cal L} \,.
\end{equation}
The energy contains a term linear in the
length~$L$ plus a correction term,
\begin{equation}
E = \sqrt{g_{tt}(u_0) g_{xx} (u_0)}\, L + K(u_0)\,,
\end{equation}
where~$K(u_0)$ can be found in~\cite{Kinar:1998vq}.  The idea now is
that the lowest part of the string is very close to the bottom of the
geometry, i.e.~$u_0 \approx 0$, so that $\sqrt{g_{tt}(u_0) g_{xx}
  (u_0)}\approx \sqrt{g_{tt}(0) g_{xx}(0)}$.  This factor then acts as
an effective string tension~$T_{\text{eff}}$.  In~\cite{Kinar:1998vq}
it was shown that if the tension is non-zero, and monotonic in~$u$
with a minimum at~$u=0$, and if all integrals converge, then~$K(u_0)$
is a correction which scales as a negative power of~$L$ or is
exponentially small in~$L$.

\begin{figure}[t]
\begin{center}
\includegraphics[width=.75\textwidth]{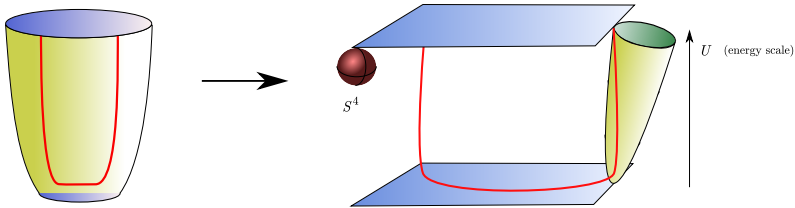}
\end{center}
\caption{The geometry of the D4-brane background: instead of
  introducing a hard cut-off in~$\text{AdS}_5$ (left), the main new
  ingredient is a circle spanned by the~$\theta$ coordinate which
  shrinks to zero size at some finite value of the radial
  direction~$u$, making the geometry everywhere smooth (right).\label{f:D4}}
\end{figure}

One way to achieve a non-zero tension and a linear potential therefore
is to ``cut off'' the AdS factor at some finite
radius~$u=u_{\text{wall}}$. This is the so-called ``hard wall
approximation'' (see
e.g.~\cite{Polchinski:2001tt,Erlich:2005qh,deTeramond:2005su}). It has
the advantage that one can compute with the relatively
simple~$\text{AdS}_5$ metric, in which many results can be obtained
analytically (see~\cite{Gursoy:2007cb,Gursoy:2007er} for the state of
the art of these and related models). However, such an artificial
cut-off of the AdS radius does not satisfy the supergravity equations
of motion, and one runs the risk that the dynamics of the gauge theory
is not correctly encoded. A more natural way to achieve a cut-off is
to make use of a supergravity solution which contains an extra
dimension which shrinks to zero size at a finite value for the radial
distance~$u$. This is similar to the way in which one can make the
boundary of a half-line smooth by turning it into a thin cigar.  Such
a situation occurs in many other D-brane metrics. An example is
\begin{equation}
\label{e:D4}
{\rm d}s^2 = \left(\frac{u}{R_{\text{D4}}} \right)^{\tfrac{3}{2}} \big[ {\eta_{\mu\nu}
  {\rm d} X^{\mu} {\rm d} X^{\nu}} + {f(u)}  {\rm d} \theta^2 \big] + 
\left(\frac{R_{\text{D4}}}{u}\right )^{\tfrac{3}{2}}\left [{\displaystyle\frac{{\rm d}u^2
  }{f(u)}} + u^2 {\rm d}\Omega_4 \right ] \,,\quad
f(u) = 1 - \left(\frac{u_\Lambda}{u}\right)^3\,.
\end{equation}
This metric is obtained by doubly Wick rotating the the near-horizon
metric of a near-extremal D4-brane~\cite{Witten:1998zw} (we use here
the coordinates of e.g.~\cite{Klebanov:1996un,Itzhaki:1998dd}). The
geometry of this space-time is depicted in figure~\ref{f:D4}. The main
feature is that the~$\theta$ circle shrinks to zero size
at~$u=u_\Lambda$, leading to a cigar-like subspace. Near the tip of
this cigar, the circle is small, and the theory is effectively
four-dimensional. The model breaks conformal invariance and has a
single mass scale.\footnote{There are various other approaches to the
  construction of confining
  backgrounds~\cite{Polchinski:2000uf,Klebanov:2000hb,Maldacena:2000yy},
  which we will not discuss here; for a review see~\cite{Aharony:2002up}.}

The minimum of~$g_{tt} g_{xx}$ now occurs at the tip of the
cigar, where the value is still non-zero. This leads to an effective
string tension given by
\begin{equation}
T_{\text{eff}}=  \frac{1}{2 \pi \alpha'}
\sqrt{-g_{tt} g_{xx}\phantom{\big|}}\Big|_{u=u_\Lambda}
= \frac{1}{2 \pi \alpha'} \left(\frac{ u_{\Lambda}}{R_{\text{D4}}} \right)^{3/2}\,.
\end{equation}%
In order to avoid a conical singularity in the $u-\tau$ plane, the
angular coordinate~$\theta$ has to be identified with period
\begin{equation}
\label{e:thetaequiv}
\theta \simeq \theta+ L_\Lambda\,,\qquad\text{with}\quad L_\Lambda = \frac{4}{3}\pi \sqrt{\frac{R_{\text{D4}}^3}{u_\Lambda}}\,.
\end{equation}
Recall that the~$\text{AdS}_5\times S^5$ case, an analysis of the
supergravity solution led to the relation $g_s N_c = R^4/l_s^4$. A
similar analysis now leads to $g_s N_c = R_{\text{D4}}^3/l_s^3$. The
five-dimensional 't~Hooft coupling (dimensionful) is given by~$\lambda_5 = g_s N_c
l_s$ and the four-dimensional one is $\lambda_4 = \lambda_5 /
L_\Lambda$, so that
\begin{equation}
\label{e:lambda4}
\lambda_4 = \frac{R_{\text{D4}}^3}{l_s^2\, L_\Lambda}\,,
\end{equation}
to be compared with~\eqref{e:lambdastring}. Recall also that all
physical quantities in the $\text{AdS}_5\times S^5$ case are expressed
in terms of the dimensionless ratio~$R^4/l_s^4$; similarly, $l_s$
disappears from physical quantities in the model discussed
here. Notice the appearance of the extra mass scale, $M_\Lambda =
2\pi/L_\Lambda$.  

\subsection{Adding flavour degrees of freedom}
\label{s:flavour}

Having obtained a confining potential, the second thing we need to do
is to introduce matter in the fundamental representation of the gauge
group. The way to achieve this was first proposed
by~\cite{Karch:2002sh}\footnote{Flavour branes were first introduced
  into confining backgrounds in~\cite{Sakai:2003wu}.}, and consists of
adding, on top of the supergravity background, a number of extra
branes. These extra branes are called ``flavour branes''. First of
all, they introduce new degrees of freedom corresponding to strings
stretching between the original branes and the new flavour
branes. These correspond to quarks in the fundamental representation
(and the number of branes thus corresponds to the number of flavours
in the gauge theory). Secondly, there are new degrees of freedom
corresponding to open strings ending with both ends on the flavour
branes. These, as we will see, correspond to bound states of quarks
and anti-quarks, i.e.~mesonic degrees of freedom. The string/gauge
correspondence then makes a prediction for the masses and couplings of
these mesons.

\begin{figure}[t]
\hspace{.1\textwidth}
\includegraphics[width=.45\textwidth]{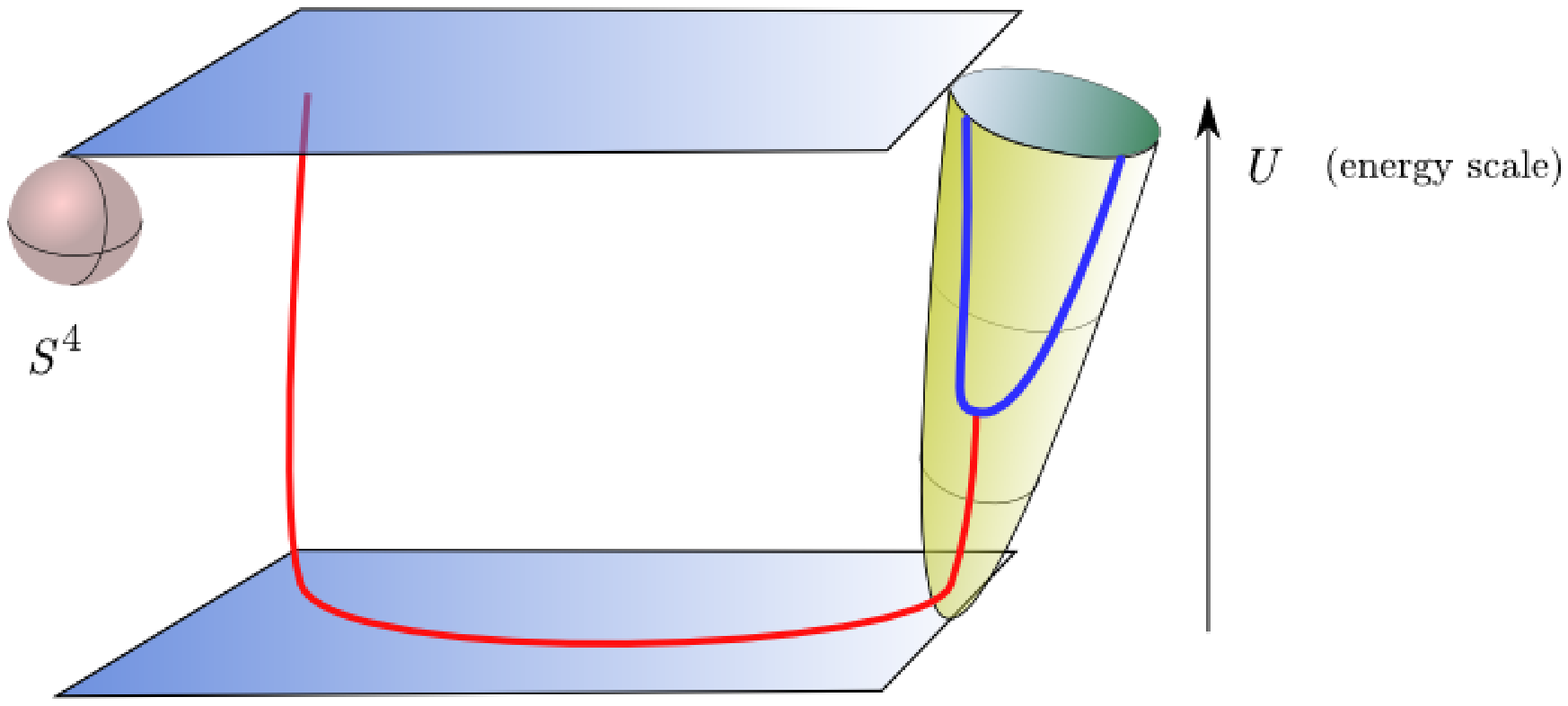}\qquad
\includegraphics[width=.45\textwidth]{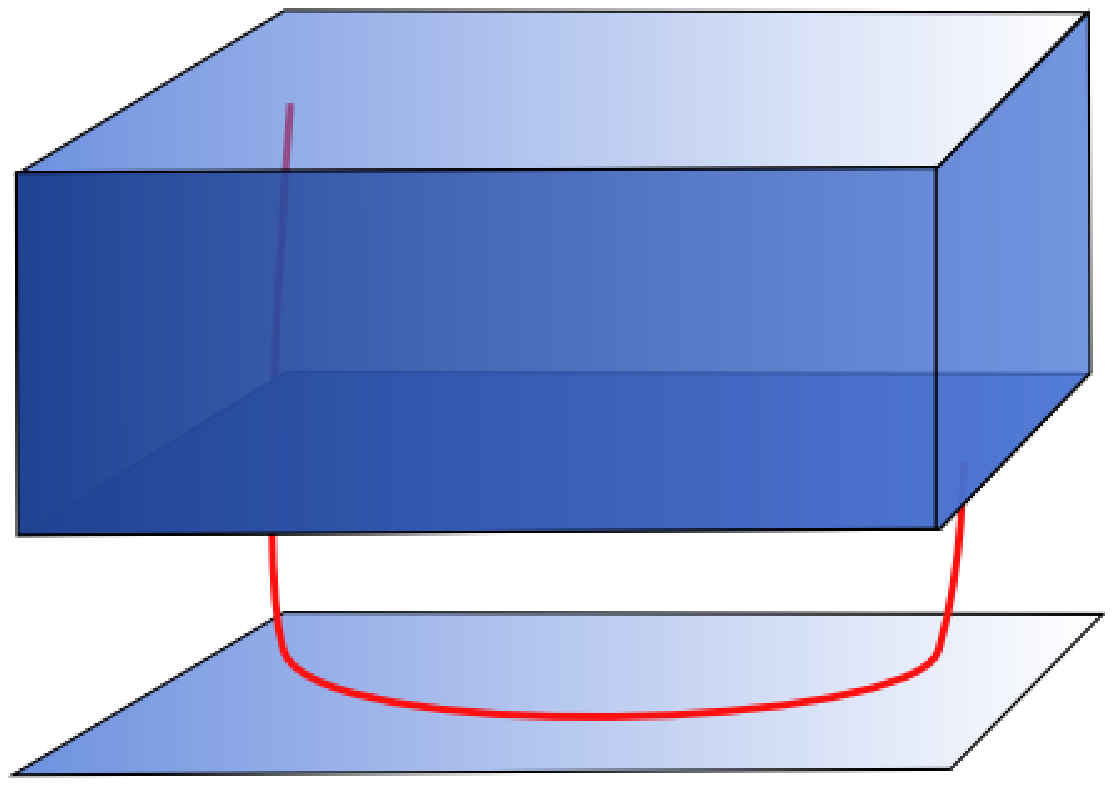}
\caption{A D8 flavour brane in the Sakai-Sugimoto model fills all of
  space except for one direction, thus leading to a curve
  $u=u(\theta)$. The picture on the right shows the same configuration
  with all directions except $x^1,x^2$ and $u$ suppressed. In this
  picture, the situation looks like a brane ``ending in thin
  air''. The long, almost rectangular curve~$u(x^1)$ depicts a~$J\gg
  1$ meson or static Wilson loop.\label{f:productspace3}}
\end{figure}

The largest dimension which a flavour brane can have while still being
non-trivial is~$8+1$, a so-called D8-brane. This leaves one function
to specify the embedding of the brane in the
background~\eqref{e:D4}. This embedding will be discussed below; it is
illustrated in figure~\ref{f:productspace3}. It turns out that the
fluctuation spectrum of this model leads to chiral
fermions~\cite{Sakai:2004cn}. However, there are of course many other
setups which one can study. The original analysis of
\cite{Karch:2002sh} employed a D7-brane added to the
$\text{AdS}_5\times S^5$ background, i.e.~a D3-D7 system. This is
still the basis of much qualitative work, because the simplicity of
the metric makes it possible to obtain many results analytically (see
e.g.~\cite{Babington:2003vm}). An alternative setup using the D4-brane
background is to introduce a D6-brane~\cite{Kruczenski:2003uq}. The
latter model exhibits an interesting phase diagram, but unfortunately
does not contain chiral fermions, so we will refrain from discussing
it here.

Let us now turn to a more quantitative analysis. The shape of the D8
flavour brane can be found by simply solving the equation of
motion. The action for the D8-brane in the background~\eqref{e:D4} is
given by
\begin{equation}
\label{e:SD8on}
S_{\text{D8}} \propto \int\!{\rm d}^4x\,{\rm d}\theta\,
u^4 \sqrt{f(u) + \left(\frac{R_{\text{D4}}}{u}\right)^3\frac{(u')^2}{f(u)}}\,,
\end{equation}
where~$u'={\rm d}u/{\rm d}\theta$.  Because the action does not depend
explicitly on the~$\theta$ coordinate there is a conserved charge
associated to translations in~$\theta$. This charge can be found by
computing the associated ``Hamiltonian'', which reads
\begin{equation}
\label{e:Ham}
{\cal H} = \frac{{\rm d}{\cal L}}{{\rm d}\theta} u' - {\cal L} =
\frac{u^4\, f(u)}{\sqrt{\displaystyle f(u) +
	 \left(\frac{R_{\text{D4}}}{u}\right)^3\frac{(u')^2}{f(u)}}} = 
u^8\, f(u)\, {\cal L}^{-1} \,.
\end{equation}
From~${\rm d}{\cal H}/{\rm d}\theta=0$ we then find for the embedding
\begin{equation}
\label{e:tausol}
\theta = u_0^4\, \sqrt{f(u_0)}\int_{u_0}^u\frac{{\rm d}u}{\displaystyle\left(\frac{u}{R_{\text{D4}}}\right)^{3/2}
  f(u)\sqrt{u^8f(u) - u_0^8 f(u_0)}} \,.
\end{equation}
Here~$u_0$ is a free integration constant, denoting the lowest point
on the~$u$ axis which is reached by the D8-brane.  The 
shape~$\theta(u)$ is depicted in figure~\ref{f:productspace3}.  Even
though this shape is thus only known implicitly and can at best be
evaluated numerically, the result~\eqref{e:tausol} is sufficient to
study the meson spectrum in this model. The~$u_0$ parameter is related
to the constituent quark mass, as we will discuss in section~\ref{s:hadfrag}.

The model is invariant under gauge transformations of the gauge field
living on the world-volume of the D8-branes. Large gauge
transformations, which asymptote to a constant independent of~$x^\mu$
as the boundaries at~$u\rightarrow\infty$ are approached, correspond
to global symmetries of the boundary theory. Because there are two
boundaries, there is a priori a~${\rm U}(N_f)\times {\rm U}(N_f)$
global symmetry, which is indeed the chiral symmetry of the
fermions. However, the gauge transformations on these two sides of the
brane have to connect smoothly at the bottom, which reduces the
freedom to only~${\rm U}(N_f)$. Hence, the connectedness of the two
halves of the D8-brane stack is the dual of chiral symmetry
breaking. 

In all models with flavour branes, there are two regimes in which we
can study mesons. One regime is that of \emph{small spin} mesons, for
spins~$J\leq 1$. These correspond to small fluctuations of the flavour
brane. The masses of these states turn out to scale just like the
glueball masses (which we have not discussed explicitly, but arise
from the fluctuation spectrum of bulk modes as we mentioned briefly in
section~\ref{s:holodict}). Explicitly, $M_{\text{small spin}} \sim
M_\Lambda$; we will discuss these mesons shortly. The other regime is that of
\emph{large spin} mesons, for~$J\gg 1$. These mesons are described by
long macroscopic strings, which have their endpoints attached to the
flavour brane. Their masses are set by the effective string tension
at~$u=u_\Lambda$, which leads to~$M_{\text{large spin}} \sim
\sqrt{\lambda} M_\Lambda$. In the supergravity regime, these states
are thus anomalously heavy compared to the small-spin ones. We will
discuss the large spin mesons in section~\ref{s:hadfrag}.

The small-spin meson spectrum contains scalar mesons, arising from the
fluctuation of the embedding coordinates of the D8-brane, as well as
vector mesons, arising from the vector fields on the brane. We can
construct an effective action for them by following the logic which we
discussed for the bulk modes in section~\ref{s:holodict}.  Let us
focus on the vector mesons. We start from the action of the D8-brane,
which is
\begin{equation}
S = V_{S^4}\int\!{\rm d}^5x\,e^{-\phi} \sqrt{ - \det\left( g_{\mu\nu} +
  2\pi\alpha'\, F_{\mu\nu}\right) } 
 = \tilde{V}_{S^4} \int\!{\rm d}^5x\,\sqrt{-g} \, F_{\mu\nu} F_{\rho\lambda}\,
  g^{\mu\rho} g^{\nu\lambda}  + \ldots
\end{equation}
(There is also a Wess-Zumino term, which however plays no role for the
determination of the meson spectrum here). Using the background
metric~\eqref{e:D4} and the embedding equation~\eqref{e:tausol}, we
get more explicitly
\begin{equation}
\label{s:explS}
S = \tilde{V}_{S^4}\int\!{\rm d}^4x\,{\rm d}u\,\Big[
u^{-1/2} \gamma^{1/2} F_{\mu\nu} F_{\mu\nu}\, 
+ u^{5/2} \gamma^{-1/2} F_{\mu u} F_{\mu u}
\Big]\,,
\end{equation}
where all indices are now contracted with the flat metric and~$\gamma
= u^8/(u^8 f(u) - u_0^8 f(u_0))$.

We now decompose the fields in terms of a four-dimensional factor and
a $u$-dependent one, as in
\begin{equation}
\begin{aligned}
F_{\mu\nu} &= \sum_{n} G_{\mu\nu}^{(n)}(x)\, \psi_{(n)}(u)\,,\\[.5ex]
F_{u\mu} &= \sum_{n} B^{(n)}_\mu(x)\, \partial_u
\psi_{(n)}(u)\,,
\end{aligned}\qquad\text{where}\qquad
B_{\mu}^{(m)}(x) = \int\!\frac{{\rm d}^4 k}{(2\pi)^4}\, e^{ik_\mu x^\mu} B_{\mu}^{(m)}(k)\,.
\end{equation}
Here~$G_{\mu\nu}^{(n)}$ is the field strength
for~$B_\mu^{(n)}$. Inserting these expansions, the polarisation
vectors can be factored out, and we end up with the following
equation of motion for the modes~$\psi_{(n)}(u)$,
\begin{equation}
\label{e:eompsi}
\int\!{\rm d}^4x\,B_\mu^{(m)} B_\mu^{(n)}\,\Big[
u^{-1/2}\gamma^{1/2} (\omega^2 - \vec{k}^{\,2} ) \psi_{(n)}
- \partial_u\Big( u^{5/2} \gamma^{-1/2}
\partial_u \psi_{(n)}\Big) \Big] = 0\,.
\end{equation}
If the equation of motion inside the brackets is satisfied for a
particular~$\omega^2 - \vec{k}^2$, and the solution is inserted back
into the action~\eqref{s:explS}, we end up with an action for massive
vector fields~$B_\mu^{(n)}$. 

Thus, in order to find the meson spectrum, we have to solve the
Sturm-Liouville problem of~\eqref{e:eompsi}. The
solutions~$\psi_{(n)}$ have to be finite on the entire D8-brane,
i.e.~on both ends of the U-shaped embedding. In order to have access to
these two sides, one introduces a new coordinate~$z$ by
\begin{equation}
u = (u_{0}^3 + u_0 z^2)^{1/3}\,.
\end{equation}
In this coordinate system, the two asymptotic ends of the D8-brane are
at $z\rightarrow +\infty$ and $z\rightarrow -\infty$ respectively.
Solving the eigenvalue problem~\eqref{e:eompsi} is possible for
generic embedding, but the equations are simplest in the case that the
D8-brane goes through the tip of the cigar, i.e.~when
$u_0=u_\Lambda$ (which makes~$\gamma = f^{-1}$). 
\begin{figure}[t]
\psfrag{m}{$m$}
\psfrag{mq}{$m_q$}
\psfrag{msq}{$m^2$}
\begin{center}
\includegraphics[width=.4\textwidth]{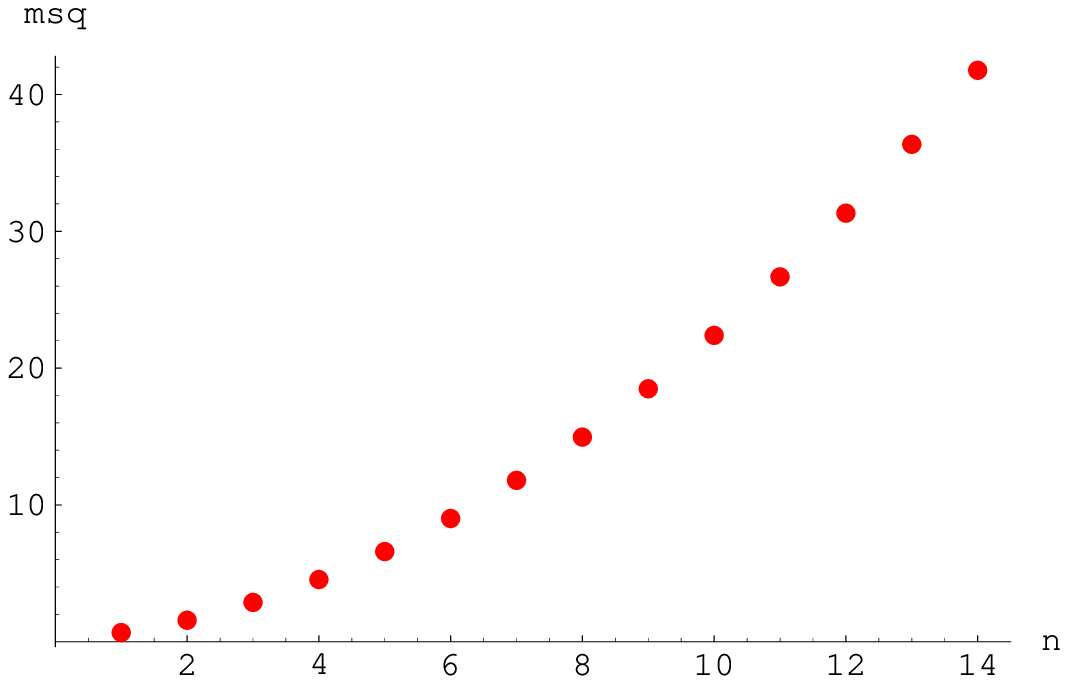}\qquad\qquad
\includegraphics[width=.4\textwidth]{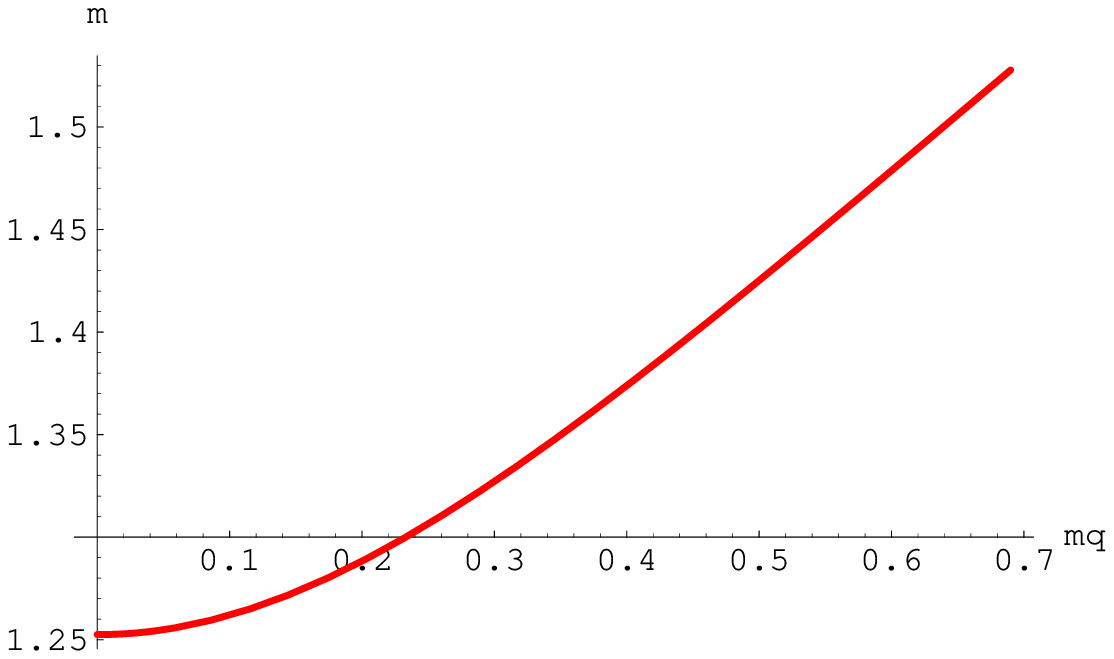}
\end{center}
\caption{The first few states of the vector meson mass spectrum of the
  Sakai-Sugimoto model (left) and the behaviour of the `$a_1$' meson
  mass as a function of the constituent quark mass (right).\label{f:SSill}}
\end{figure}

Let us first look at the massless sector. We will see that this
contains pions described by a chiral Lagrangian. In the~$z$
coordinate, there are two zero-mass modes (cf.~\eqref{e:eompsi}), being
combinations of a normalisable and a non-normalisable one near the two
boundaries,
\begin{equation}
\psi_+ = \frac{1}{2}(1 \pm \psi_0)\,,\qquad
\psi_0 = \frac{2}{\pi}\arctan z \,.
\end{equation}
For the expansion we use (in the~$A_z=0$ gauge)
\begin{equation}
A_{\mu} = i \xi_+ \partial_\mu \xi_+^{-1}(x)\,\, \psi_+(z) + i \xi_- \partial_\mu
\xi_-^{-1}(x)\,\, \psi_-(z)\,,
\end{equation}
where we will leave~$\xi_+$ and $\xi_-$ arbitrary for the time being;
the presence of two modes reflects that there is still a residual
gauge invariance present.  For the holographic derivation of the
chiral Lagrangian for pions, we focus on the~$F_{\mu z}^2$ term, do
one partial integration in the~$z$ direction and thus write the action
as
\begin{equation}
\label{e:Srewrite}
\begin{aligned}
S &= \text{EOM} + \int\!{\rm d}^4x\, \Tr\big( A_\mu\, F_{\mu z} \big)\, (1+z^2)
\Big|_{z=-\infty}^{\infty}\\[1ex]
  &= \text{EOM}  + \int\!{\rm d}^4x \!\!\!\!\begin{aligned}[t]
 &\Tr\Big[ \xi_+ \partial_\mu \xi_+^{-1}\,\, \psi_+ 
\Big( \xi_+ \partial_\mu \xi_+^{-1}\,\,\psi_+' +
      \xi_- \partial_\mu \xi_-^{-1}\,\,\psi_-' \Big) 
  (1+z^2)\Big]_{z=+\infty}\\[1ex]
  - &
 \Tr\Big[ \xi_- \partial_\mu \xi_-^{-1}\,\, \psi_-
\Big( \xi_+ \partial_\mu \xi_+^{-1}\,\,\psi_+' +
      \xi_- \partial_\mu \xi_-^{-1}\,\,\psi_-' \Big) 
  (1+z^2)\Big]_{z=-\infty} \,.
\end{aligned}
\end{aligned}
\end{equation}
Using the explicit expression for the modes~$\psi_\pm$ the boundary
terms reduce to
\begin{equation}
\label{e:bdyremain}
S = \frac{1}{\pi}\int\!{\rm d}^4x\, \Tr\Big[
\big(\xi_+\partial_\mu \xi_+^{-1} - \xi_-\partial_\mu \xi_-^{-1}\big)^2\Big]\,,
\end{equation}
From this result we should now read off how to identify the~$\xi_\pm$
parameters with the pion \mbox{$U= \exp i\pi$}.  There are many
choices which one can make, but one useful one is \mbox{$\xi_-=1$} and
\mbox{$\xi_+ = U$} (note that for this particular gauge choice, the
contribution from the \mbox{$z=-\infty$} boundary is absent
altogether). This leaves the chiral action
\begin{equation}
\label{e:SkyrmefromSS}
 S = -\frac{1}{\pi} \int\!{\rm d}^4x\, \Tr\Big[ \partial_\mu U
 \partial_\mu U^{-1}\Big]\,.
\end{equation}
By also taking into account the commutator terms in the non-abelian
field strength, one in fact recovers the full Skyrme
action for the pions~\cite{Sakai:2004cn}. More importantly, the string
dual also predicts how the Skyrme model pions should be coupled to an
infinite tower of massive vector mesons.

Let us therefore continue with the massive meson modes. It is no
longer possible to solve~\eqref{e:eompsi} analytically, but it is
possible to obtain the solutions numerically. A Frobenius analysis shows that
the asymptotic solution of~\eqref{e:eompsi} behaves, up to an overall
normalisation, as
\begin{equation}
\psi(z) \sim \frac{1}{z} \big( 1 + \frac{c_1}{z^{2/3}} +
\frac{c_2}{z^{4/3}} + \ldots\big)\,,
\end{equation}
where the first few coefficients~$c_i$ are given by
\begin{equation}
c_1 = -\frac{9}{10\,u_0} (\omega^2 - \vec{k}^2)\,,\qquad
c_2 = \frac{81}{280\,u_0^2} (\omega^2 -\vec{k}^2)^2\,.
\end{equation}
Using this as an initial condition at~$z=-\infty$, solving the
differential equation numerically and demanding normalisability
at~$z=+\infty$ then yields a discrete spectrum of masses, starting
with~\cite{Sakai:2004cn}
\begin{equation}
(\omega^2 - \vec{k}^2)^{CP} = 0.67^{--},\,\,\, 1.57^{++},\,\,\,2.87^{--},\,\,\, 4.55^{++},\ldots\,,
\end{equation}
in units of the single mass scale of the model.  The parity and charge
conjugation properties are also indicated.  By generalising this
analysis for arbitrary~$u_0$ one can also compute the dependence of
the masses on the quark masses~$m_q$, see figure~\ref{f:SSill} (this
relies on the relation between~$u_0-u_\Lambda$ and the constituent
quark mass, which we will discuss in the next section).

It is tempting to compare these results to actual meson masses, and
one finds that the ratios are in reasonable agreement with
experiment. A similar agreement is found when the analysis is
generalised to interactions of mesons~\cite{Sakai:2004cn}. The goal
here is not so much to argue that the string/gauge duality can pin
down these numbers accurately (which it cannot), but rather to show
that the model seems to capture some of the universal features which
also make many other models get the numbers right. The exciting aspect
is that this patterns follows from equations which at first sight have
nothing to do with mesons. \footnote{To be fair, there are also
  several qualitative aspects which the Sakai-Sugimoto model does not
  get right. It has an exotic~$0^{++}$ state, but lacks $J^{+-}$
  states altogether.  The mass of the pions (identically zero) is not
  related to the constituent quark masses (see
  however~\cite{Casero:2007ae} for directions on how to cure this). The
  Kaluza-Klein modes on the sphere lead to a tower of modes which are
  not separated from the others by a large mass gap. As in all weakly
  curved models, there is a large gap between mesons of spin~$\geq 1$
  and those of higher spin. Finally, because the supersymmetry
  breaking scale is of the order of the meson mass scale, there is a
  whole tower of fermionic mesons which are partners of the bosonic
  ones; these are obviously also not present in QCD.}

\subsection{Long strings and hadron fragmentation}
\label{s:hadfrag}

\begin{figure}[t]
\begin{center}
\includegraphics[width=.7\textwidth]{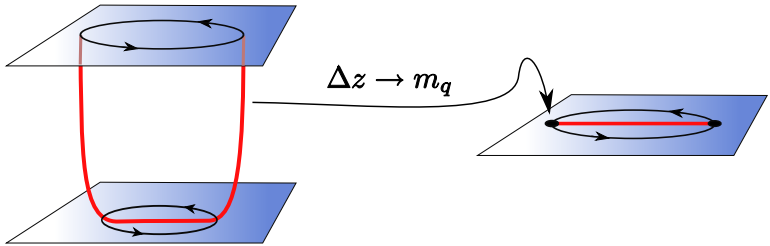}
\end{center}
\caption{A large-spin meson in the string/gauge theory
  description. The endpoints of the string map to the quark and
  anti-quark in the gauge theory picture, while the string between
  them extends deeply into the curved geometry. The masses of the
  quarks are now represented by the long vertical string
  segments~\cite{Kruczenski:2004me}.\label{f:projection}}
\end{figure}

Let us now turn to large-spin mesons. These are described by long,
U-shaped strings which hang from the flavour branes, with their
end-points attached to it, as in figure~\ref{f:projection}. One can
solve the equation of motion for such a string, and find the energy
and angular momentum as a function of the separation~$L$ and the
angular velocity~$\omega$,
\begin{equation}
\begin{aligned}
 E &= 2 \frac{T_{\text{eff}}}{\omega} \left( \arcsin (\omega L) +
   \sqrt{\frac{m_q}{T_{\text{eff}} L}} \right)\,,\\[1ex]
J &= \frac{T_{\text{eff}}}{\omega^2} \left( \arcsin (\omega L) +
  (\omega L)^2 \,
  \sqrt{\frac{m_q}{T_{\text{eff}} L}} \right)\,.
\end{aligned}
\end{equation}
Intriguingly, by eliminating~$\omega$, one finds at leading order
precisely the relation between energy and angular momentum for a
relativistic four-dimensional string with massive
endpoints~\cite{Kruczenski:2004me}. The masses of the endpoints map to
the masses of the vertical string segments. These stretch between the
bottom of the flavour brane at~$u=u_0$ and the bottom of the geometry
at~$u=u_\Lambda$, and thus
\begin{equation}
m_q = \frac{1}{2\pi\alpha'} \int_{u_\Lambda}^{u_0} \sqrt{-g_{tt}
  g_{uu}} {\rm d}u\,.
\end{equation}
For large~$u_0$, and using the metric~\eqref{e:D4}, this shows a
simple linear relation between the constituent quark mass and the
parameter~$u_0$.

In the real world, large-spin mesons are extremely unstable (the
Particle Data Book lists decay rates only for spins to about~6; higher
spin mesons decay too rapidly~\cite{PDBook}). However, the situation
is quite different in the large-$N_c$ limit. As we have seen, this
limit implies an arbitrarily small string coupling constant, and decay
processes are suppressed by at least one power of~$N_c$. Therefore, it
makes sense to study decaying large strings in the string/gauge theory
duality.

Moreover, there is reasonable evidence for the existence of
short-lived large-spin mesons in a more indirect form. In any particle
collider, after one leaves the hard, perturbative regime, one enters a
stage in which the quarks and leptons that have been produced combine
and form hadrons. This hadronisation stage is not understood from
first principles, but several Monte-Carlo tools for the analysis of
collider data are based on phenomenological string breaking models
such as the Lund
model~\cite{Sjostrand:1982fn,Andersson:1983ia}. Although the
large-spin mesons are too short-lived to be seen directly, one does
observe the decay products; the fact that programs such as PYTHIA
reproduce collider data so far quite well lends support to the idea
that large-spin mesons really do exist.

One of the more popular phenomenological models for large-spin meson
decays is the CNN model~\cite{Casher:1978wy}. It is based on the idea
of modelling the force between the quarks in a meson by a
chromoelectric flux tube. The probability of such a meson decaying is
then directly related to the probability of creating a new
quark/anti-quark pair inside the flux tube. This probability is the
standard Schwinger pair production probability, given by the
characteristic exponential suppression factor
\begin{equation}
P_{\text{pp}} = \exp\left({-\frac{\pi}{2} \frac{m_q^2}{T_{\text{eff}}}}\right)
\quad\rightarrow\quad P = \frac{T_{\text{eff}}^2}{\pi^3} \sum_{n=1}^\infty \frac{1}{n^2} 
  \exp\left({-\frac{\pi}{2} \frac{n\, m_q^2}{T_{\text{eff}}}}\right)
\end{equation}
The decay width is the given by multiplying this with the volume of
the flux tube,
\begin{equation}
\label{e:CNNres}
\Gamma = L\cdot \pi r_t^2\, P
= L \cdot T_{\text{eff}}\, (\text{number})\, \sum_{n=1}^\infty \frac{1}{n^2} 
  \exp\left({-\frac{\pi}{2} \frac{n\, m_q^2}{T_{\text{eff}}}}\right)
\end{equation}
For vanishing quark masses, we find that the decay width divided by
the total mass is a constant, because the mass of the flux tube is
also linear in $L$,
\begin{equation}
m_q = 0\quad\rightarrow\quad
M = T\cdot L\quad\rightarrow\quad
\frac{\Gamma}{M}=\text{const}.
\end{equation}
When the quark masses are non-zero we have, for a fixed total mass,
a shorter flux tube,
\begin{equation}
\frac{L}{M}=\frac{2}{\pi\,T_{\text{eff}}}- \frac{m_1+m_2}{2 T_{\text{eff}} M} 
+ {\cal O}\left(\frac{m_i^2}{M^2}\right)\,.
\end{equation}
This results in~$\Gamma/M$ versus~$M$ plots not being straight
lines~\cite{Gupta:1994tx}. 

\begin{figure}[t]
\begin{center}
\includegraphics[width=.9\textwidth]{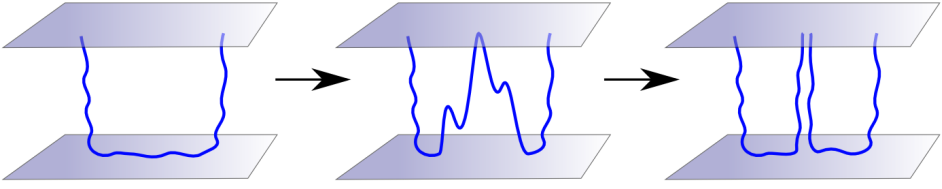}
\end{center}
\caption{A fluctuating string may touch a flavour brane, at which
  point it can split and reconnect, thus resulting in two open strings.\label{f:fluct}}
\end{figure}

The string/gauge theory correspondence claims to give a more fundamental
link between mesons and long strings. The question is thus whether the
results of the CNN model can be reproduced from the string/gauge
theory correspondence. The first thing to note is that fundamental
quarks are no longer present, but their masses are a reflection of the
masses of the vertical segments of the U-shaped string. Instead of
pair-creating quarks, we should thus consider processes which lead to
new vertical segments.

Now the U-shaped string is a quantum object, and this implies that it
fluctuates. In fact, it is possible for the horizontal part of the
string to fluctuate up to a flavour brane, as in
figure~\ref{f:fluct}. If that happens, the point at which the string
touches the flavour brane can split and the two new endpoints can
reconnect to the brane. This leads to two U-shaped strings, and
therefore our original meson has decayed into two new mesons.

The probability for this process to happen can be calculated, at least
in the approximation where the gravitational field is approximately
constant between the bottom of the space-time (the ``infrared wall'')
and the flavour brane. The splitting probability is a product,
\begin{equation}
{\cal P} = {{\cal P}_{\text{fluctuate}}} \,\times\, {{\cal P}_{\text{split}}}\,.
\end{equation}
The first factor describes the probability for the string to fluctuate
to the flavour brane, while the second factor describes the
probability that it will actually split there. The second factor has
to do with a process which takes place on the flavour brane, where the
curvature is constant. Therefore, we can use standard flat-space
results for this. The first factor, on the other hand, requires
precise control over the wave function of the string.

\begin{figure}[t]
\hspace{1ex}
\psfrag{P}{\hspace{-2.5em}\raisebox{2ex}{\smaller $\Gamma/(T_{\text{eff}} {\cal 
P}_{\text{split}} L)$}}
\psfrag{zb}{\smaller $\displaystyle\frac{z_B}{\sqrt{\alpha'_{\text{eff}}}}$}
\includegraphics[width=.4\textwidth]{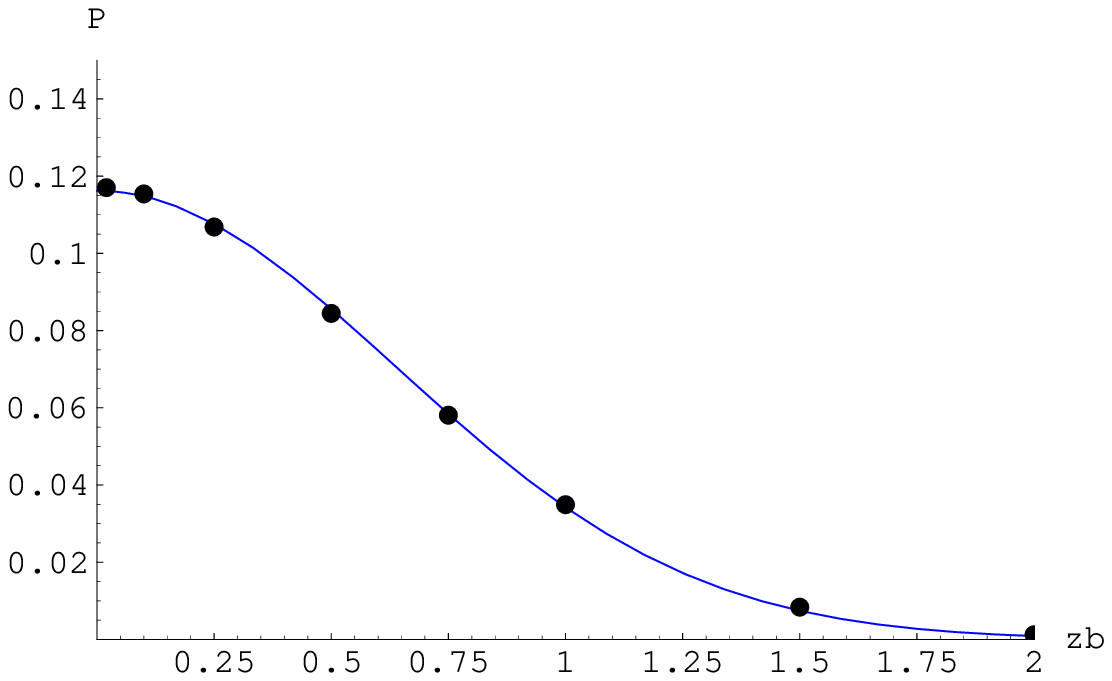}\qquad\qquad
\psfrag{K892}{\small \raisebox{-2ex}{$K^*(892)$}}
\psfrag{K1430}{\small \raisebox{-1.5ex}{$K^*_2(1430)$}}
\psfrag{K1780}{\small\hspace{-6em} $K^*_3(1780)$}
\psfrag{K2045}{\small $K^*_4(2045)$}
\psfrag{K2380}{\small $K^*_5(2380)$}
\psfrag{GdM}{\small $\Gamma/M$}
\psfrag{M}{\small $M~(\text{GeV})$}
\includegraphics[width=.4\textwidth]{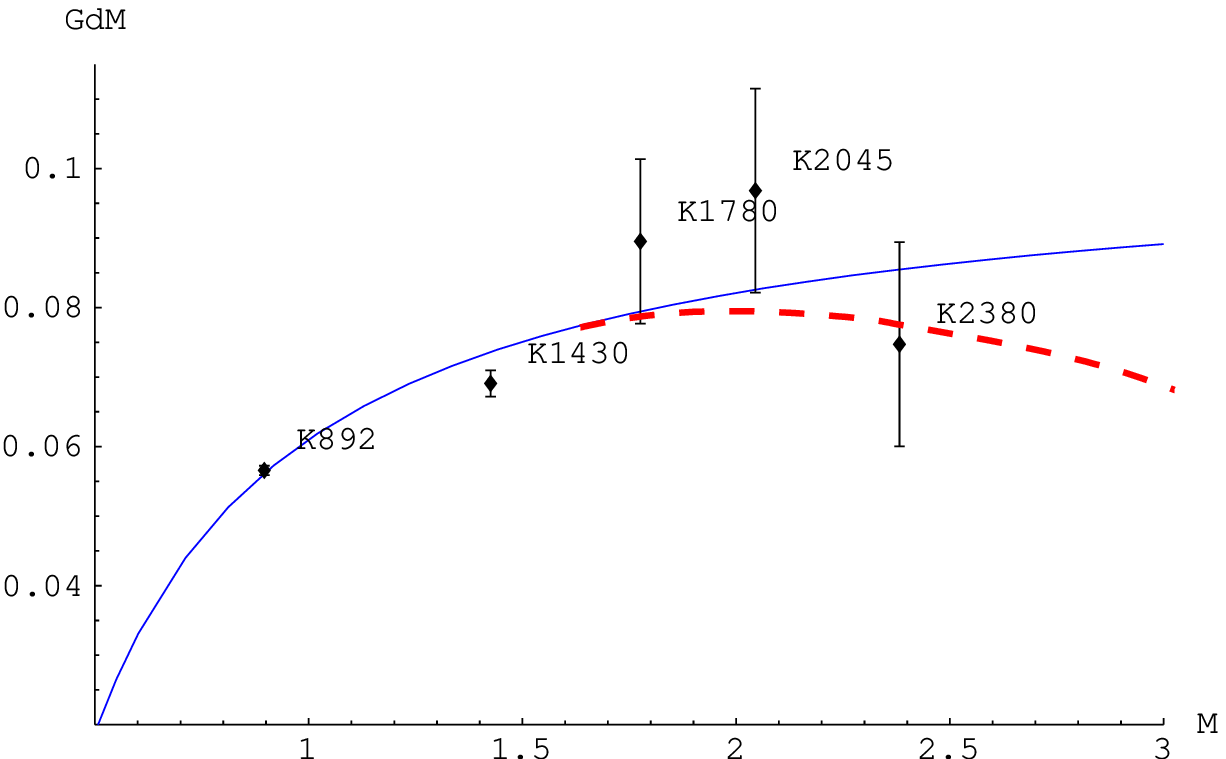}\qquad\qquad
\vspace{2ex}
\caption{The decay width per unit length for a meson computed using a
  holographic dual (left). Corrections from the curvature of the
  background generically predict that the decay widths are suppressed
  for very large spins, which is symbolically depicted in the plot
  for the $K^*$ decay widths on the
  right (the continuous, blue curve is the leading order result on which the CNN
  model and the holographic computation agree).\label{f:fluctresults}}
\end{figure}

Recall how this is done in flat space. The Polyakov action for the string in
flat space is given by
\begin{equation}
\label{e:SflatNM}
S = \int\!{\rm d}\tau {\rm d}\sigma\Big( \dot{X}^2 - {X'}^2 \Big)\qquad\rightarrow\qquad
S = \int\!{\rm d}\tau \sum_{n} \Big(\dot{a}_n^2 - \omega_n^2 a_n^2\Big)\,,
\end{equation}
where we have decomposed in Fourier modes. The wave function is a
function over the variables~$a_n$, i.e.
\begin{equation}
\Psi\big[ \{a_n\} \big] = \prod_n {\Psi_n}\big[ a_n(X) \big]\,,
\end{equation}
where all the~$\Psi_n$ are harmonic oscillator wave functions. If we
put all oscillators in the ground state, we end up with
\begin{equation}
\Psi_n\big[ a_n \big] \sim \exp\Big( - \omega_n a_n^2 \Big)\,.
\end{equation}
In curved space, the logic is the same, only the mode expansions
become more complicated. One has to find fluctuations around the
U-shaped string which are such that the action decouples as a sum of
normal modes, with coefficients~$\eta_n$, like in~\eqref{e:SflatNM}. The upshot is that the wave
function now takes the form
\begin{equation}
\Psi\big[\{\eta_n\} \big] = \prod_n \Psi_n\big[\eta_n (X)\big] \, .
\end{equation}
The goal is then to compute the total probability that the string
touches the flavour brane, i.e.~compute the integral 
\begin{equation}
\label{norm-prob}
\Gamma = T_{\text{eff}} {\cal P}_{\text{split}}\,\times\,
   \int'_{\{ \eta_n\}} \big|\, \Psi\big[\{\eta_n\}\big] \, \big|^2
   \, K\big[\{\eta_n\}\big]  \, ,
\end{equation}
where we integrate over all configurations such that the string
touches the flavour brane with at least one point.  The
factor~$K\big[\{\eta_n\}\big]$ is a measure factor with the
dimension of length. It measures, for a given string configuration,
the size of the segment(s) of the string which intersect(s) the
flavour brane. This is a tricky computation because of the boundary
conditions. However, various estimates and a computation using a
string bit model agree on the result~\cite{Peeters:2005fq}. Namely,
one indeed finds that the decay width is close to a Gaussian in the
distance to the flavour brane, or in terms of gauge theory variables,
a Gaussian in~$m_q$ (see figure~\ref{f:fluctresults}). One finds
precisely the CNN result~\eqref{e:CNNres}.

Interestingly, one can go beyond the approximation of flat space and
try to take into account the curvature of the background. This should
encode possible corrections to the Lund decay widths. For the model at
hand, the wave functions can be computed analytically in terms of
Mathieu functions; the upshot is that the wave function for the~$n$-th
mode becomes
\begin{equation}
\label{effpsi}
\Psi[\eta_n] \sim \exp\left[ - \frac{2}{3\alpha'} (R_{\text{D4}}^3u_\Lambda)^{1/2} 
    \big(a_n(b/2) + b\big)^{1/2}\, \eta_n^2 \right]\quad\text{where}\quad
b = \frac{27}{4}\pi^2 \frac{J}{g^2_{\text{YM}}
  N_c} +\ldots\,,
\end{equation}
where~$a_n(b)$ are the Mathieu characteristic functions which at
leading order in $b$ behave as~$n^2$ ($b$ is given in the
approximation where the quarks are light). In this model, and various
related ones, we thus conclude that the generic trend is for curvature effects to
suppress the decay of mesons with very large
spin~\cite{Peeters:2005fq}.

\section{Finite temperature and phase transitions}
\subsection{Temperature and Euclidean geometry}

\begin{figure}[t]
\begin{center}
\includegraphics[width=.7\textwidth]{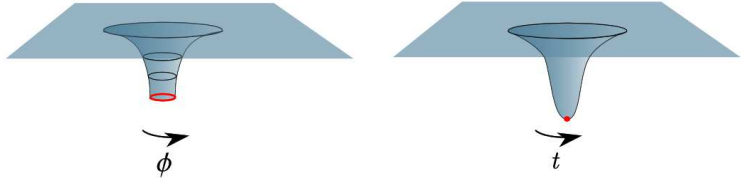}
\end{center}
\caption{Lorentzian versus Euclidean sections of the Schwarzschild
  geometry. In the Euclidean section, the circle spanned by the~$t$
  coordinate shrinks to zero size at the horizon, leaving a conical
  singularity unless~$t\sim t+2\pi \cdot 4m$.\label{f:EuclidSS}}
\end{figure}

Much of the interest in the string/gauge theory correspondence comes
from applications to finite-temperature physics.  Before we can
discuss string duals to gauge theories at finite temperature, let us
first recall how black holes are related to finite temperature. The
simplest way to see this, and the way which connects most
straightforwardly to what we will say later, is to look at the
Euclidean section of a black hole geometry. If we start from the
Schwarzschild geometry and perform a Wick rotation~$t = i\tau$, we end
up with the geometry
\begin{equation}
{\rm d}s^2 = \left(1-\frac{2m}{r}\right) {\rm d}t^2 +
\left(1-\frac{2m}{r}\right)^{-1} {\rm d}r^2 + r^2 {\rm d}\Omega^2\,.
\end{equation}
If we now introduce a new coordinate~$\tilde{r} = r-2m$ and then~$\rho^2
= 8m\tilde{r}$, we end up very close to the horizon~$r=2m$ with the
metric
\begin{equation}
{\rm d}s^2 = \frac{\tilde{r}}{2m} {\rm d}t^2 +
\frac{2m}{\tilde{r}}{\rm d}\tilde{r}^2 + 4m^2\,{\rm d}\Omega^2
= \left(\frac{\rho}{4 m}\right)^2 {\rm d}t^2 + {\rm d}\rho^2 + 4m^2\,{\rm d}\Omega^2\,.
\end{equation}
In the~$\rho,t$ subspace, this is precisely a flat metric in polar
coordinates, with~$t$ playing the role of the angular coordinate. We
thus see that we naturally get a compactified Euclidean time. In order
to identify the period, note that unless we identify~$t\sim
t+2\pi\cdot 4m$, this metric will exhibit a conical
singularity. Therefore, we choose this particular size of the
Euclidean circle, and end up with a temperature
\begin{equation}
T = \frac{1}{8\pi m}\,.
\end{equation}
This is precisely the temperature that can also be obtained by using
Hawking's original derivation of the temperature of radiation produced
by a black hole.  The difference between the Lorentzian and Euclidean
black hole geometry is once more visualised in
figure~\ref{f:EuclidSS}.

Let us now look at the thermal characteristics of the string
geometries that play a role in the string/gauge theory
correspondence. The first metric which we encountered was the one of
$\text{AdS}_5$. After a Wick rotation, it can be written in the form
\begin{equation}
\label{e:AdS}
{\rm d}s^2_{\text{AdS}_5} = 
\left(1+\frac{r^2}{R^2}\right){\rm d}t^2 + 
\left(1+\frac{r^2}{R^2}\right)^{-1}{\rm d}r^2 + r^2\,{\rm d}\Omega^2_3\,.
\end{equation}
This metric is regular everywhere for any radius of the time-like
circle, i.e.~$t\sim t+\beta$ for any arbitrary $\beta$. However, there
exists another solution to the supergravity equations of motion, with
the same asymptotic geometry~$S^1\times S^3$, whose Euclidean section takes the form
\begin{equation}
\label{e:AdSBH}
{\rm d}s^2_{\text{AdS-BH}_5} = 
\left(1+\frac{r^2}{R^2}- \frac{\mu}{r^2}\right){\rm d}t^2 + 
\left(1+\frac{r^2}{R^2}- \frac{\mu}{r^2}\right)^{-1}{\rm d}r^2 + r^2\,{\rm d}\Omega^2_3\,.
\end{equation}
This is the so-called $\text{AdS}_5$ black hole, and this geometry has
a horizon (in the Lorentzian section) at
\begin{equation}
r^2 = r_+^2 = (R/2)(-R+\sqrt{R^2+4\mu^2})\,.
\end{equation}
Moreover, with an identification~$t\sim t+\beta$ the geometry has a
conical singularity unless
\begin{equation}
\label{e:betarplus}
\beta = \frac{4\pi R^2 r_+}{4 r_+^2 + 2 R^2}\,.
\end{equation}
This fixes the temperature of the black hole.

The two geometries~\eqref{e:AdS} and~\eqref{e:AdSBH} compete with each
other in the Euclidean partition sum,
\begin{equation}
Z =   \sum_{\text{configurations}}\!\!\! \exp[ - \beta F ]\,.
\end{equation}
where~$F$ is the free energy, related to the Euclidean action by~$S_E = \beta F$.
The dominant one is the one with smallest Euclidean action. One can
compute this action explicitly~\cite{Witten:1998zw}, but the details are not
important. The phase diagram of this system is rather simple because
there is only one parameter (the temperature) to vary. One finds that
below a certain critical temperature, the dominant configuration
is~\eqref{e:AdS}, while above this temperature, the partition sum is
dominated by the black hole~\eqref{e:AdSBH}.\footnote{To be precise,
  there are always two black hole geometries competing in the
  partition sum, since at a fixed temperature there are two solutions of~\eqref{e:betarplus}
  for~$r_+$. The one which dominates at large temperature is the one
  with the largest value of~$r_+$.}

\subsection{Phase diagrams}

The phase diagram for the~$N=4$ theory is we discussed above is not
particularly realistic, as it does not exhibit confinement and also
does not contain any quark degrees of freedom. We can use it as a
guideline, however, to study more realistic gauge theories, such as
the confining model discussed in
section~\ref{s:coulomb_to_confining}. If we Wick rotate the
metric~\eqref{e:D4} we end up with
\begin{equation}
\label{e:EuclidD4}
{\rm d}s^2 = \left(\frac{u}{R_{\text{D4}}} \right)^{3/2} \big[
  {\rm d}t^2
  + \delta_{ij}{\rm d}x^i {\rm d}x^j + f(u) {\rm d} \theta^2 \big] + 
\left(\frac{R_{\text{D4}}}{u}\right )^{3/2}\left [\displaystyle\frac{{\rm d}u^2
  }{f(u)} + u^2 {\rm d}\Omega_4 \right ] 
\end{equation}
One feature which sets this metric apart from the Euclidean ones we
have seen so far is that it contains \emph{two} circles. One is the
circle spanned by the~$\theta$ coordinate, with period~$L_\Lambda$
(see~\eqref{e:thetaequiv}), which was also present in the Lorentzian
geometry. The other one is the new Euclidean time-like circle spanned
by~$t$.

\begin{figure}[t]
\begin{center}
\includegraphics[width=.9\textwidth]{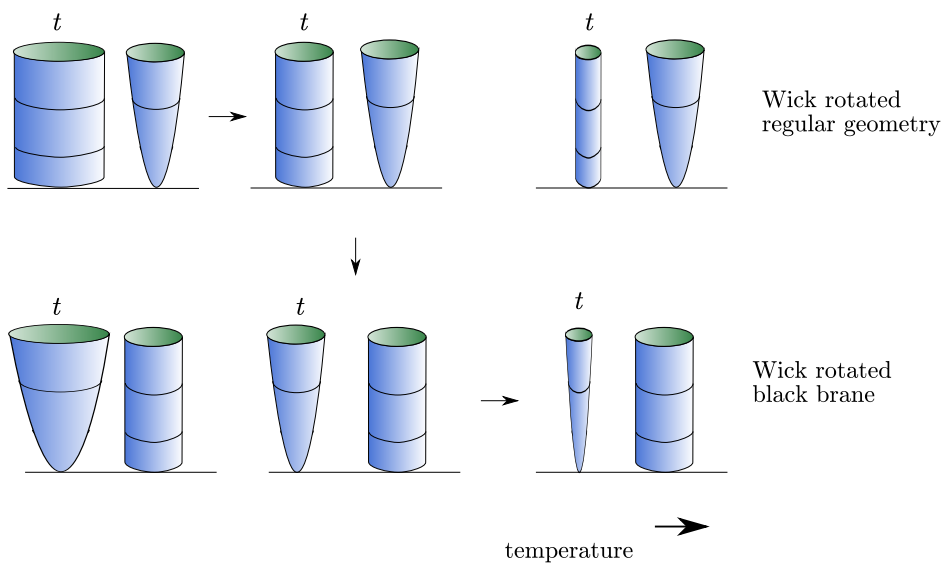}
\end{center}
\vspace{-2ex}
\caption{The two D4-brane based geometries which compete in the
  Euclidean partition sum: both are described by the same
  metric~\eqref{e:EuclidD4}, the difference is only which circle
  changes size as we change the temperature. For the regular geometry,
  temperature is related to the size of the cylinder, while for the
  black hole geometry, temperature is related to the radius of the
  cigar.\label{f:euclidean}}
\end{figure}

\begin{figure}[t]
\begin{center}
\includegraphics[width=.8\textwidth]{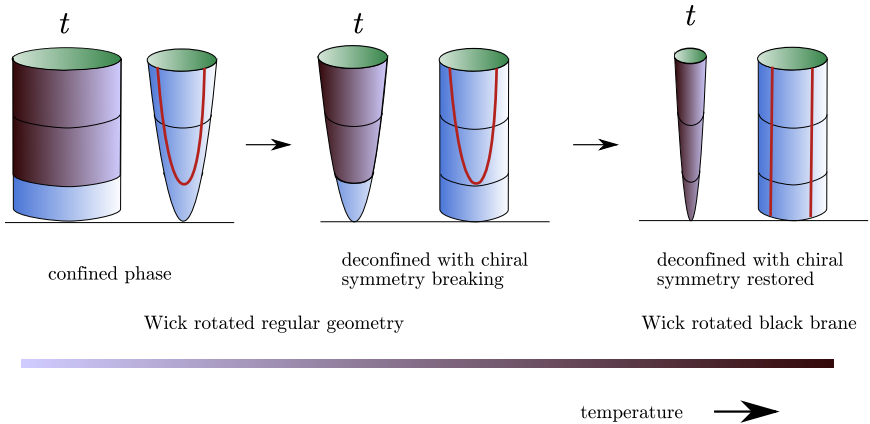}
\end{center}
\vspace{-2ex}
\caption{The three possible phases of the Sakai-Sugimoto model,
  showing the two background geometries (regular or black) and
  the two possible brane embeddings (connected or disconnected).\label{f:threephases}}
\end{figure}

We could therefore wonder whether we could perhaps view~$\theta$ as
the thermal circle? If we Wick rotate back, but now on the~$\theta$
coordinate, we end up with a metric which near~$u=u_\Lambda$ takes the
form
\begin{equation}
\label{e:intermediateTmetric}
- f(u) {\rm d}\theta^2 + (\ldots){\rm d}u^2 + \delta_{ij}{\rm d}x^i{\rm d}x^j\,.
\end{equation}
Recalling that~$f(u)=1-u_\Lambda^3/u^3$, this metric is very much like
a black hole metric, with a horizon at~$u=u_\Lambda$ and a temperature
given by $T = L_{\Lambda}^{-1}$.  We thus have access to two different
thermal interpretations of the Euclidean metric~\eqref{e:EuclidD4},
one based on a regular Lorentzian geometry in which~$t$ is the
Euclidean time, and one based on a Lorentzian black hole (or rather
``black brane'') geometry in
which~$\theta$ is the Euclidean time.

In the Euclidean partition sum for quantum gravity, these two
configurations again compete, and the one with smallest action will
dominate at a given temperature. This balance is sketched in
figure~\ref{f:euclidean}. At low temperatures, the dominating geometry
turns out~\cite{Aharony:2006da} to be the regular one (with a large
radius cylinder factor and a fixed-radius cigar). As we decrease the
radius of the cylinder, we encounter a point where the radius is equal
to that of the cigar. At this point, the Wick rotated black hole phase
takes over. This transition is first-order because the solutions do
not smoothly connect, and continue to exist as separate solutions both
below and above the transition. The spectrum of the glueball
fluctuations in the phase described by~\eqref{e:intermediateTmetric}
is \emph{continuous}, signalling the deconfinement of the gluonic
degrees of freedom.

An order parameter of this phase transition is the Polyakov loop,
corresponding to a string wrapped around the time direction. In the
low-temperature phase the time circle is non-contractible and not the
boundary of a disc, so that the Polyakov loop vanishes. After the
transition, this circle becomes contractible, resulting in a non-zero
expectation value for the Polyakov loop. Other order parameters are
the Wilson loop (which has a linear quark/anti-quark potential in the
first background but vanishing tension in the second) and the
behaviour of the free energy as a function of~$N_c$ (namely~$\sim
N_c^0$ for the first background and~$\sim N_c^2$ for the second), and
the presence of a glueball spectrum with a mass gap.

Just as at zero temperature (discussed in section~\ref{s:flavour}),
we can now add flavour D8-branes to the system. This is depicted in
figure~\ref{f:threephases}. The D8-brane always wraps the Euclidean
time circle, and describes a curve in the subspace described by the
other circle and the $u$~direction. At low temperatures, the branes
are connected, leading to chiral symmetry breaking as discussed
in section~\ref{s:flavour}. Once in the black hole phase, however, there comes a
temperature at which it is energetically more favourable for the
D8-branes to disconnect and go vertically down to the horizon. Chiral
symmetry is restored here.

\begin{figure}[t]
\begin{center}
\includegraphics[width=.6\textwidth]{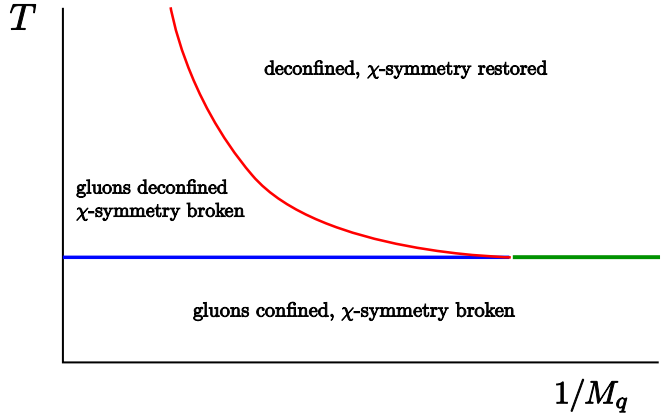}
\end{center}
\caption{The phase diagram for the Sakai-Sugimoto
  model~\cite{Aharony:2006da}. For large enough constituent quark
  masses, one encounters three phases as the temperature is increased:
  starting with a fully confined phase, first the gluons deconfine,
  and later the chiral symmetry gets restored and mesons deconfine
  too. For smaller quark masses the deconfinement goes in one step.\label{f:asy_phases}}
\end{figure}

Whether or not the intermediate phase, which has deconfined gluons but
confined mesons, exists depends on the masses of the constituent
quarks. The phase diagram of the model is given in
figure~\ref{f:asy_phases}~\cite{Aharony:2006da}.

\subsection{Mesons and meson melting}

In the intermediate-temperature phase, in which there is a black brane
background encoding the dynamics of deconfined gluons, there may still
be bound states of the matter degrees of freedom. In order to analyse
these, one needs to analyse the dynamics of the flavour brane in the
black brane background. Just as before, there are low-spin massive
vector mesons described by the small fluctuations of the gauge fields
on the flavour brane. The thermal (pole) masses, defined for
homogeneous modes by
\begin{equation}
\label{e:massdef}
\partial_0^2 B^{(n)}_i = - m_n^2 B^{(n)}_i \, , 
\end{equation}
can be computed by solving a fluctuation equation similar to the
zero-temperature case displayed in~\eqref{e:eompsi}. It now reads
\begin{equation}
\label{e:eompsi2}
-u^{1/2} \gamma^{-1/2}\,{f(u)^{1/2}}\,\partial_u\left( u^{5/2}
 \gamma^{-1/2} {{f(u)^{1/2}}}\partial_u \psi_{(n)}\right) = R_{\text{D4}}^3 \, m_n^2\, \psi_{(n)}\,.
\end{equation}
The only difference is the insertion of two extra
factors~$\sqrt{f(u)}$. The spectrum remains discrete, labelled by
the mode number~$n$ and parity. By keeping the asymptotic geometry fixed,
including the separation~$L$ between the flavour branes, and changing
the temperature, one can determine the behaviour of the masses as a
function of temperature. 

\begin{figure}[t]
\begin{center}
\psfrag{rho}{``$\rho$''}
\psfrag{a1}{``$a_1$''}
\psfrag{m2}{$m^2$}
\psfrag{T}{$T/T_c$}
\psfrag{m2}{$m^2$}
\psfrag{uT}{$T/T_c$}
\psfrag{m2}{$m^2$}
\psfrag{E2}{$E^2$}
\psfrag{T}{$\;T/T_c$}
\hspace{-2em}\includegraphics[width=.42\textwidth]{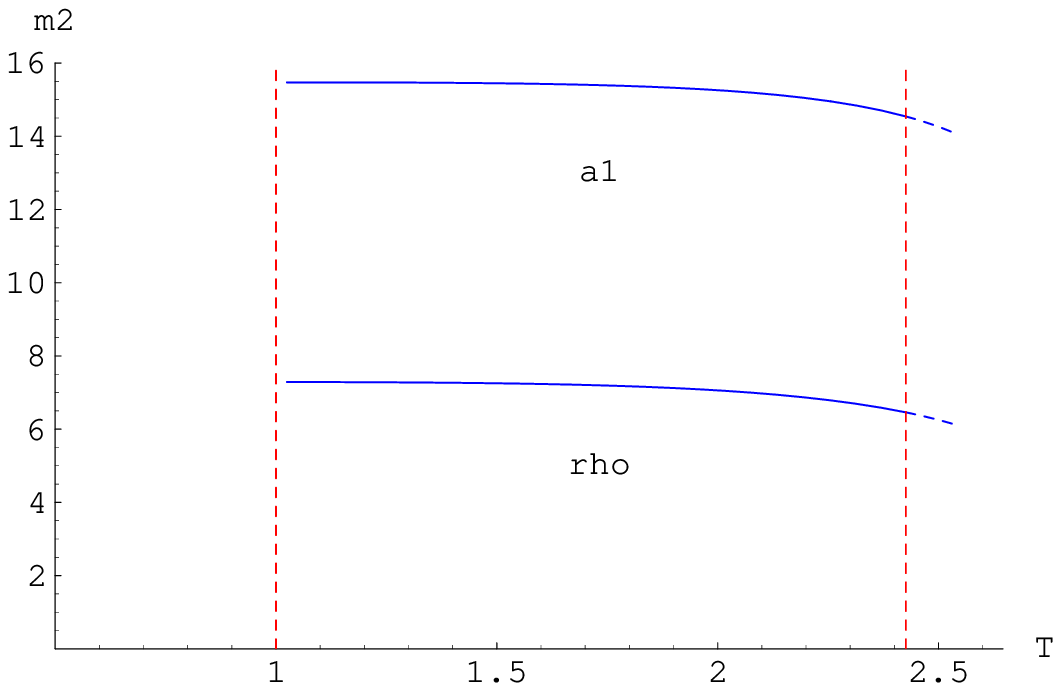}\qquad
\includegraphics[width=.42\textwidth]{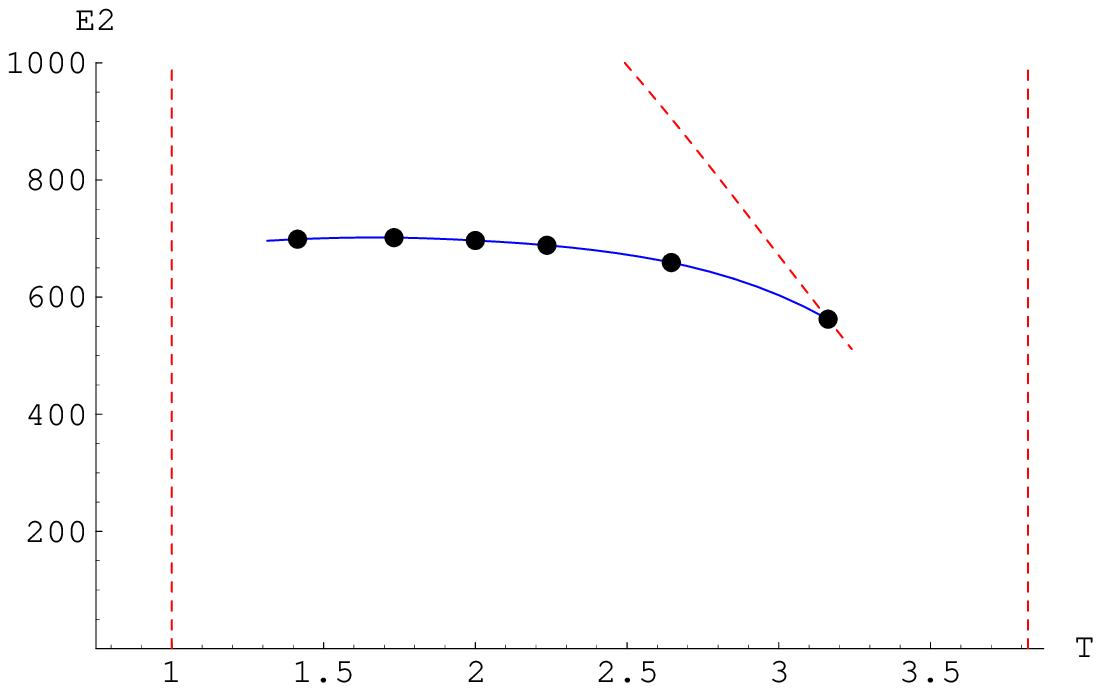}
\end{center}
\caption{The behaviour of the masses of small-spin mesons (left) and
  large-spin mesons (right) as a function of
  temperature. While small-spin mesons survive all the way up to the
  transition into the high-temperature phase, large-spin mesons cease
  to exist beyond a certain temperature, when they get too close to the
  horizon and disappear from the spectrum.\label{f:thermalmesons}}
\end{figure}

By comparing with the zero-temperature result, one observes that the
masses of light mesons decrease as the temperature is
increased~\cite{Peeters:2006iu}. The temperature dependence of the
masses of the ``$\rho$'' and ``$a_1$'' mesons are shown on the left
panel of figure~\ref{f:thermalmesons}. These mesons survive all the
way up to the transition into the phase where chiral symmetry is
restored. Large-spin mesons, on the other hand, can be shown to
disappear from the spectrum once they reach a certain critical size;
see the right panel of figure~\ref{f:thermalmesons}.

These findings are in qualitative agreement with lattice
computations~\cite{Karsch:1999vy,Gottlieb:1996ae}, although the
Sakai-Sugimoto model does not exhibit the degeneracy of the~$\rho$
and~$a_1$ meson masses at high temperature. Moreover, we will see
below that the spectrum becomes continuous in the phase where chiral
symmetry is restored. Therefore, the phase transition from
intermediate to high temperature, which is first-order in our model,
is rather different from the one in QCD. However, what we see is
consistent with the fact that mesons in the chirally symmetric phase
are extremely unstable in the large-$N_c$ limit: their decay widths
scale as a positive power of~$N_c$.

Finally, let us briefly discuss the high-temperature phase. In this phase, the
minimal free energy of the system is attained when the two stacks of
branes are disconnected (see the third panel of
figure~\ref{f:threephases}). The profile of the left and right stacks
of branes is characterised by $u'\equiv {\rm d}u/{\rm d}\theta\rightarrow
\infty$. The differential equation for the modes, analogous
to~\eqref{e:eompsi} and~\eqref{e:eompsi2} for the other two phases, is now
\begin{equation}
\label{e:eqhighT}
-u^{1/2}\,f(u)^{1/2}\,\partial_u\left( u^{5/2}
 f(u)^{1/2}\partial_u \psi_{(n)}\right) = R_{\text{D4}}^3 \, m_n^2\, \psi_{(n)}\,.
\end{equation}
which differs from the one at intermediate temperature~\eqref{e:eompsi2} by
the absence of the~$\gamma$ factors. 

In this phase, the massless pion disappears from the spectrum. The
reason is that this mode, which would be given by~$\phi^{(0)} =
u^{-5/2} f(u)^{-1/2}$, is no longer normalisable. The norm compatible
with the differential equation~\eqref{e:eqhighT} leads, for the
would-be pion mode, to the divergent integral
\begin{equation}
\int_{u_T}^\infty\!{\rm d}u\, u^{5/2} f(u)^{-1/2} \Big| u^{-5/2} f(u)^{-1/2}\Big|^2\,.
\end{equation}
Although this integral is convergent at the upper boundary, it
diverges at the lower boundary because \mbox{$f(u) \sim \sqrt{u-u_T}$}
for $u\sim u_T$. The crucial ingredient is that the D8-branes extend
all the way down to the horizon at~$u=u_T$, which did not happen in
the intermediate temperature phase. Thus, in accordance with the fact
that chiral symmetry is restored in the high-temperature phase, we see
that the Goldstone boson has disappeared. The remainder of the
spectrum turns out to be continuous.

\section{Using strings to study the quark-gluon fluid}

Given that string duals to gauge theories are computationally
tractable only when the 't~Hooft coupling is large, it is only natural
that new observed gauge theory phenomena at strong coupling would
catch the interest of string theorists. The discovery of the strongly
coupled quark gluon fluid at RHIC thus came as a nice surprise not
only for experimentalists~\cite{Shuryak:2004cy}. 

One of the first properties of the fluid which caught the attention
is its low viscosity~\cite{Teaney:2003kp}. Using the string theory
dual of gluon fluids, this turned out to have an extremely natural
interpretation in terms of properties of black hole horizons. An
excellent review article of this topic has meanwhile
appeared~\cite{Son:2007vk}, so we will refrain from discussing this
issue here in detail. Instead, we will focus on two other applications
which are of a more recent nature, namely the study of drag forces and
the study of screening lengths. These both involve the dynamics of
quarks inside the deconfined gluon fluid, and require an intermediate
temperature phase as we have seen in previous sections.

\subsection{Drag forces and jet quenching}

When a quark/anti-quark pair is produced inside a particle collider,
the quark and anti-quark will typically end up hadronising into
back-to-back jets. However, when pair production takes place inside
the quark-gluon fluid, and the pair is produced close to the boundary
of the fluid ball, the situation is different. One of the quarks will
be able to escape the fluid and form a jet as usual, but the other
quark has to travel a long way to the other side of the fluid and
will lose energy in this process. The result is a signal with one
well-defined jet and a much smeared-out signal on the other side of
the detector. This phenomenon, known as ``jet quenching'', thus tells
us something about the interaction of the fluid with the quarks.
From Bremsstrahlung-type calculations one can estimate the energy loss
which quarks experience at weak coupling. At strong coupling, the drag
force can be computed using the string/gauge theory
correspondence~\cite{Herzog:2006gh,Gubser:2006bz}.

The main ingredient is that a single quark can be modelled by a string
ending on a flavour brane and hanging down all the way to the
horizon. We thus use a stationary ansatz for the string of the form
\begin{equation}
x(u,t) = v\, t + x(u)\,.
\end{equation}
The computation was originally done for an $\text{AdS}_5$ black hole
with flat horizon, but can easily be extended to more complicated
metrics. For an ansatz of the form above, the action of the string
reduces to
\begin{equation}
\label{e:danglingstring}
S_{\text{string}} = \int\!{\rm d}\tau{\rm d}\sigma\, \sqrt{-g} 
= \int\!{\rm d}\tau{\rm d}\sigma\,
\sqrt{ - g_{tt} g_{xx} (x')^2 - g_{tt} g_{uu} - g_{xx} g_{uu} v^2}\,.
\end{equation}
Since there is no explicit~$x$-dependence, the equation of motion is
given by a simple ordinary differential equation, which can be
integrated once trivially; making use of~\eqref{e:danglingstring} we find
\begin{equation}
\label{e:xprime}
\frac{{\rm d}}{{\rm d}u}\left(
\frac{g_{tt} g_{xx}}{\sqrt{-g}} \frac{{\rm d}}{{\rm d}u} x(u)
\right) = 0\qquad\rightarrow\qquad
(x(u)')^2 = \frac{C^2\,( -g_{tt} g_{uu} - g_{xx} g_{uu} v^2)}{g_{tt}
    g_{xx} ( g_{tt} g_{xx} + C^2)}\,.
\end{equation}
As we have seen twice in a different situation, we will not need more
than the derivative~$x'$ in order to find the drag force.
Substituting the result for~$x'$ back into the action, the result is
\begin{equation}
S_{\text{string}} = \int\!{\rm d}\tau{\rm d}\sigma\,
\sqrt{ -\frac{g_{tt} g_{uu} g_{xx} ( g_{tt} + g_{xx} v^2 )}{C^2 +
    g_{tt} g_{xx}}}\,.
\end{equation}
This expression tells us how to determine the constant~$C$. If we do
not choose it appropriately, the string will reach a point~$u=u_c$
where the numerator inside the square root will change sign. This
happens at~$g_{tt} + g_{xx} v^2=0$, which is a point \emph{above} the
horizon, which is located at~$g_{tt}=0$. Clearly, a physical solution cannot have
imaginary action. Therefore, a solution which reaches~$u=u_c$ will
have to have its $C$ chosen such that the denominator changes sign
here too. After~$C$ has been fixed in this way, the shape of the string can
always be found by integrating~\eqref{e:xprime} once more, if
necessary numerically. 

However, in order to determine the drag force which has to be exerted
on the string, we do not need to know the shape~$x(u)$. We merely have
to compute the momentum density
\begin{equation}
P^1{}_x := \frac{\delta S}{\delta x(u)'} = -\frac{1}{2\pi\alpha'}\frac{g_{tt} g_{xx} x(u)'}{\sqrt{-g}}\,.
\end{equation}
The force at the endpoint is related to this by~$F = {\rm d}p_x/{\rm
  d}t = \sqrt{-g} P^1_x$.
For the Sakai-Sugimoto model, this calculation leads
to~\cite{Burikham:2007kp}
\begin{equation}
F_{\text{drag}} = -\frac{1}{2\pi l_s^2} \frac{v}{\sqrt{1-v^2}}
\left(\frac{u_\Lambda}{R_{\text{D4}}}\right)^{3/2} = 
- \frac{1}{2\pi} \frac{v}{\sqrt{1-v^2}} \left(\frac{4\pi}{3}\right)^3
 \lambda_4\,T^2\,,
\end{equation}
where use was made of the expression for the temperature in terms of
the circle radius, $T = L_\Lambda^{-1}$ together
with~\eqref{e:thetaequiv} and the expression for the four-dimensional
't~Hooft coupling~\eqref{e:lambda4}. Note that there is no explicit
power of~$N_c^2$ here, so this expression does not scale as the
entropy density~$s$ or as a power of the energy
density~$\epsilon^{3/4}$. The result is also independent of the quark
mass. If required this drag force can be used to compute the ``jet
quenching parameter''~\cite{Herzog:2006gh}.

These calculations can of course be repeated for other non-conformal
gravity duals~\cite{Buchel:2006bv} as well as for backgrounds modified
by $\alpha'$ corrections~\cite{Armesto:2006zv}, with qualitatively
similar results. In contrast to the entropy bound, however, the drag
force is not universal for all gauge theories which have a gravity
dual (neither the numerical coefficients nor the behaviour as a
function of the 't~Hooft coupling are the same for all gauge theories
with string duals).

\subsection{Screening length}

The last topic we will discuss is the computation of the screening
length of the colour force in a strongly-coupled gluon fluid. From
lattice calculations, it is now known that the quark/anti-quark
potential starts to flatten out when the separation exceeds a certain
temperature-de\-pen\-dent scale~$L_s$. By dimensional reasoning one
expects~$L_s\sim 1/T$, and explicit computations fix the
proportionality constant; see e.g.~\cite{Kaczmarek:2005ui}. 

In order to apply this result to dynamical situations occuring in the
strongly-coupled quark gluon fluid, one would need to know the
behaviour of the screening length when the quark/anti-quark pair is
not at rest with respect to the medium, but has a non-zero
velocity. This is much harder to do on the lattice as it involves
time-dependence at finite temperature. However, as the drag force
calculation shows, the string/gauge theory correspondence has the
potential to deal with such time-dependent finite temperature
situations efficiently.

The situation we need to study is that of a large-spin meson, for
which we want to know the maximum possible size as a function of the
temperature and the velocity with respect to the medium. In the
Sakai-Sugimoto background these rotating mesons can be obtained
numerically~\cite{Peeters:2006iu} both at zero and at finite
temperature. These colour-singlet states do not experience a drag
force like the isolated quarks do, because the string stretching
between the quarks never reaches the horizon. From the field-theoretic point of
view this makes sense, as the absence of a colour monopole moment
means that these states do not couple directly to gluon degrees of
freedom. Because of the absence of drag, mesons do not experience any
energy loss when propagating through the quark-gluon fluid: no force
is necessary to keep them moving with fixed velocity. However, their
shape is certainly velocity dependent.  For higher velocity, the
distance to the horizon decreases, leading to a lower melting
temperature.

\begin{figure*}[t]
\begin{center}
\psfrag{E2}{$E^2$}
\psfrag{J}{$J$}
\psfrag{w}{$\omega$}
\vspace{2ex}
\includegraphics[width=.3\textwidth]{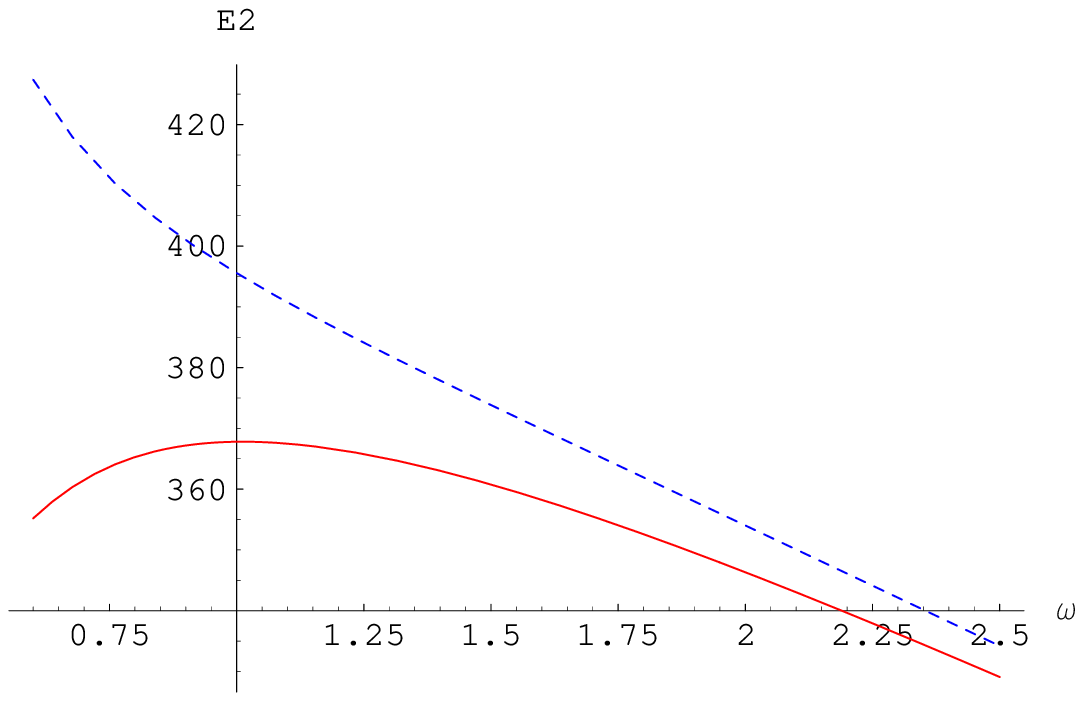}\quad
\includegraphics[width=.3\textwidth]{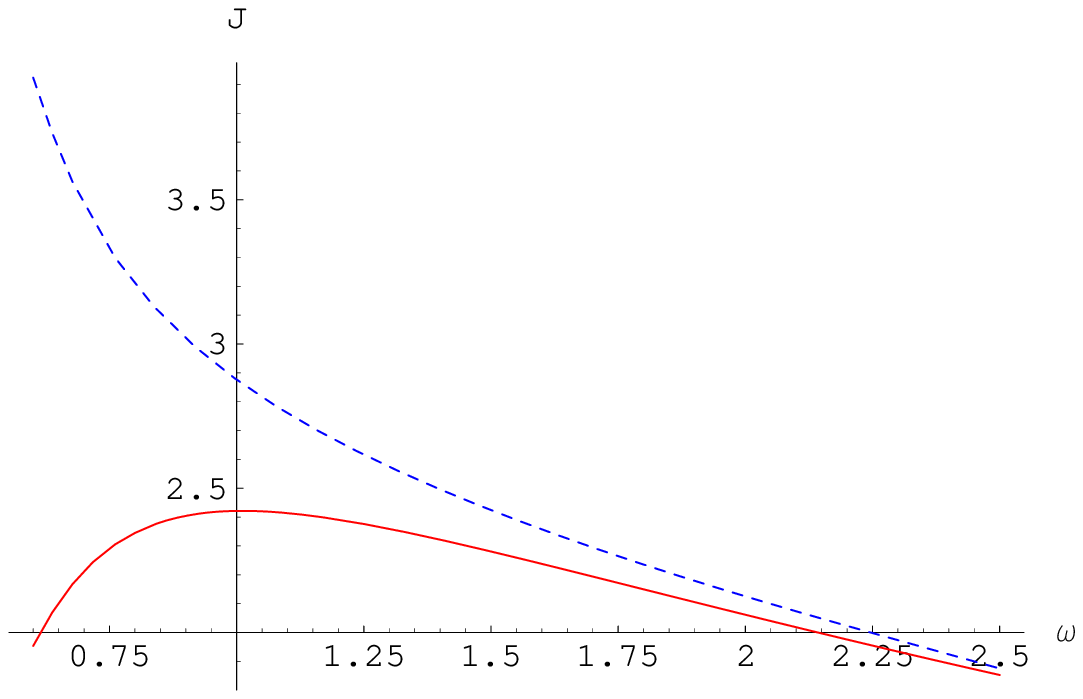}\quad
\includegraphics[width=.3\textwidth]{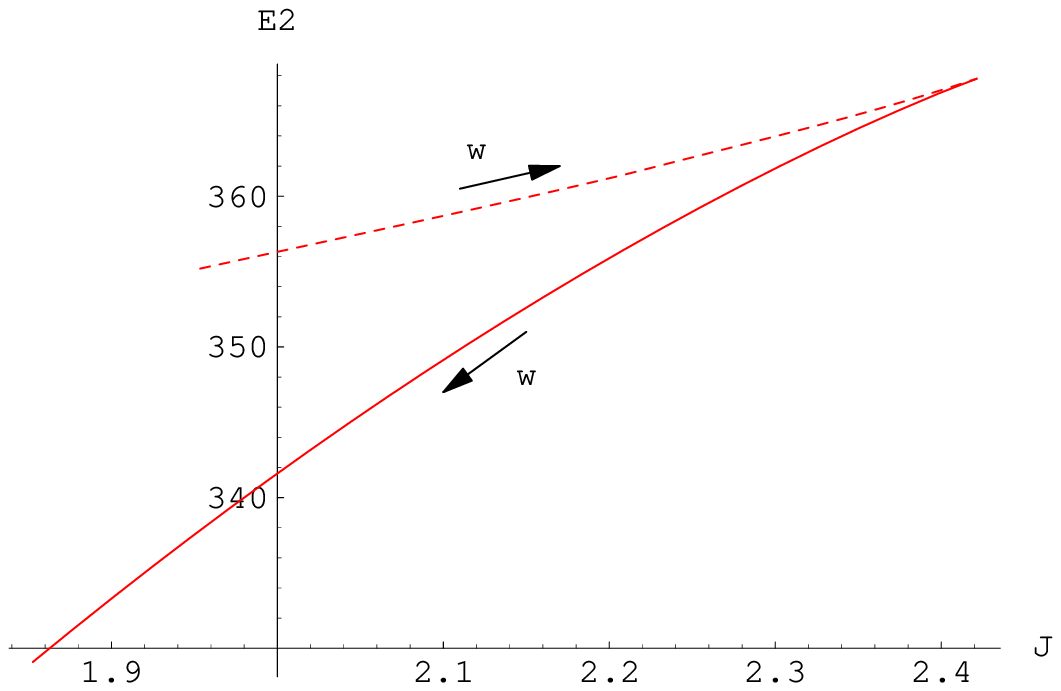}
\end{center}
\caption{The energy (squared) and angular momentum of U-shaped
strings, as a function of the frequency~$\omega$. The dashed (blue)
curves are for zero temperature, while the solid (red) curves are for
the intermediate-temperature regime.\label{f:EJom}}
\end{figure*}

To make this more precise, let us plot the energy and angular momentum
as a function of the angular velocity~$\omega$ (see
figure~\ref{f:EJom}).  When we eliminate the variable~$\omega$ from
the left and middle plots in figure~\ref{f:EJom}, we obtain the plot
on the right, expressing the energy versus the spin. The appearance of
a maximum in energy and spin is clearly visible. We also see that the
two states with identical spin~$J$ actually have different energy: the
ones with smaller~$\omega$ are more energetic than the ones with
larger~$\omega$. Therefore, the upper branch to the left of the maxima
in the plots of figure~\ref{f:EJom} is presumably unstable and will
decay to the lower branch.

\begin{figure}[t]
\vspace{2ex}
\begin{center}
\psfrag{J}{$J$}
\psfrag{omega}{$\omega$}
\psfrag{v0}{\small $v_y=0$}
\psfrag{v98}{\small $v_y=0.98$}
\psfrag{v09}{\small $v_y=0.9$}
\psfrag{rho}{$\rho$}
\psfrag{u}{$u$}
\includegraphics[width=.45\textwidth]{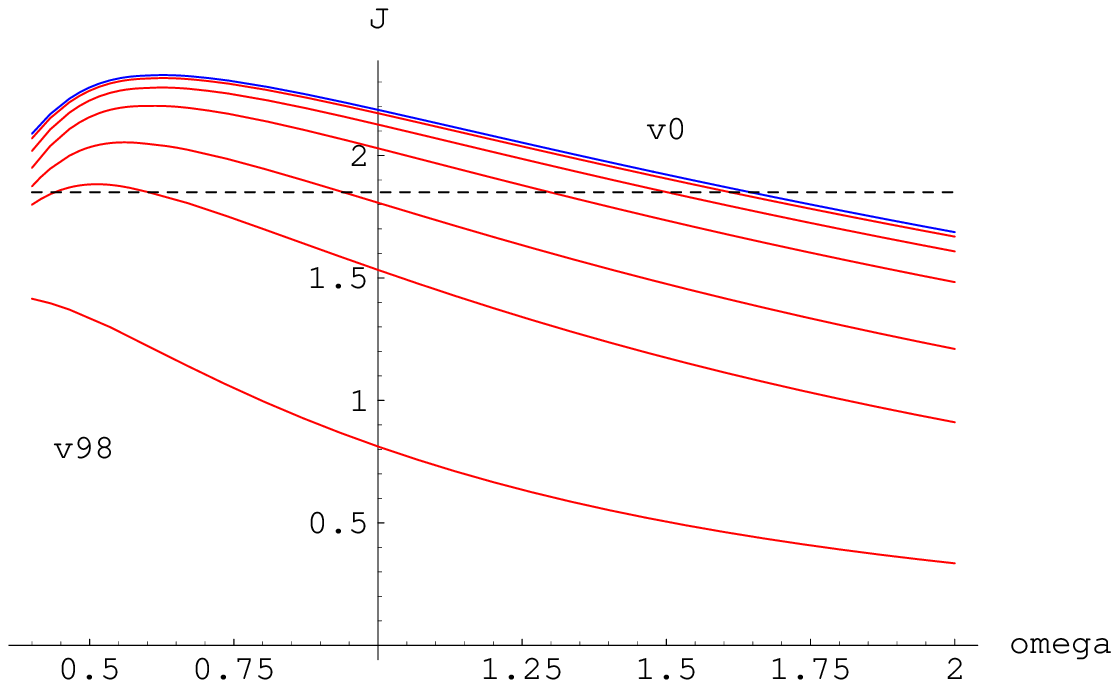}\quad
\psfrag{LatJmax}{$L^{4d}$}
\psfrag{v}{$v$}
\includegraphics[width=.45\textwidth]{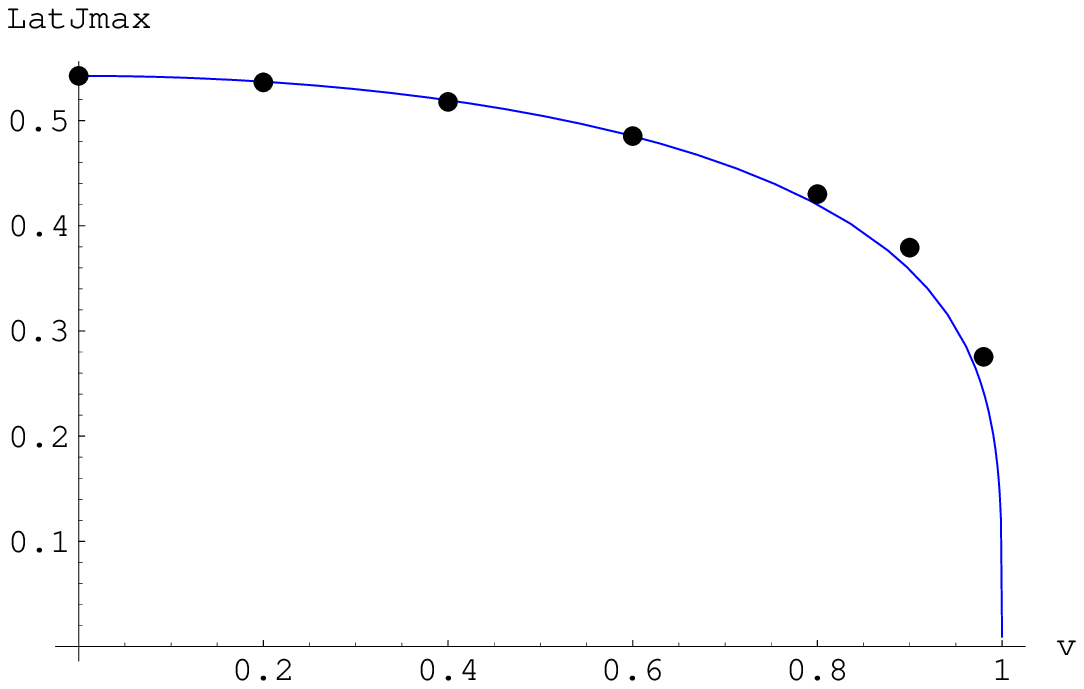}
\end{center}
\caption{The behaviour of the spin~$J$ as a function of the angular
  and transverse velocities~$\omega$ and $v_y$ respectively
  (left). The size of the maximum-spin meson depends on~$v_y$. This
  leads to a plot of the four-dimensional size $L^{4d}$ for
  maximum-spin mesons as a function of the transverse velocity
  (right). The blue curve depicts the relation $L^{4d}(v=0) \cdot
  (1-v^2)^{1/4}$, as obtained analytically for Wilson loops
  in~\cite{Liu:2006nn}.\label{f:final}}
\end{figure}

These curves can be analysed also for various values of the transverse
velocity~$v$. This yields the left plot in figure~\ref{f:final}. From
this plot we see that for a fixed temperature there is a
\emph{maximal} value of the spin which a meson can carry. It is
natural to interpret the temperature at which this happens as the
critical temperature at which a meson of spin~$J_{\text{max}}$ melts.
Thus we conclude that (as intuitively expected) the dissociation
temperature of large-spin mesons is \emph{spin dependent}.  As the
temperature increases, the maximal value of the spin that a meson can
carry decreases, i.e.~for given quark mass, higher-spin mesons melt at
lower temperature. There is thus a critical velocity beyond which a
meson of fixed spin has to dissociate. Similarly, the four-dimensional
size of the meson decreases with increasing velocity. The data is
approximated rather well~\cite{Peeters:2006iu} by the relation
\begin{equation}
\label{e:LatJmax}
  L^{4d}_{\text{max-spin}}(v) \approx L^{4d}_{\text{max-spin}}(v=0)\cdot (1-v^2)^{1/4}\,.
\end{equation}
This fit was motivated by the analytic results of~\cite{Liu:2006nn},
in which a similar dependence on the velocity was found for the
screening length, or more precisely, the maximum interquark distance
for Wilson loops in an AdS black hole background. See also related
results in~\cite{Chernicoff:2006hi}.

\section{Conclusions}

We have reviewed how the ``new'' string description of gauge theories
has led to several intriguing results about strong-coupling gauge
theory dynamics, with interesting connections to RHIC
physics. Although we still lack a proper proof of the string/gauge
theory correspondence (even in the prototype AdS/CFT case), it has
certainly already led to a fascinating geometrical point-of-view
on many problems which are hard to tackle with lattice or effective
field theory models. It will be interesting to see how this new
approach continues to complement other theoretical tools in the
future. 

\vfill
\section*{Acknowledgements}

We thank Ofer Aharony and Cobi Sonnenschein for an inspiring
collaboration on many of the topics discussed in this review and for
feedback on the first draft. We would also like to thank the organisers
of the 45th Winter School on Theoretical Physics in Schladming for the
invitation to present these lectures, and the audience for many
interesting comments and discussions. The work of KP was partially
supported by VIDI grant 016.069.313 from the Dutch Organisation for
Scientific Research (NWO).  \eject


\begin{thebibliography}{60}
\expandafter\ifx\csname natexlab\endcsname\relax\def\natexlab#1{#1}\fi

\bibitem[Maldacena(1998)]{mald2}
J.~M. Maldacena, ``The {large-$N$} limit of superconformal field theories and
  supergravity'', {\em Adv.\ Theor.\ Math.\ Phys.} {\bf 2} (1998) 231--252,
\href{http://xxx.lanl.gov/abs/hep-th/9711200}{{\tt hep-th/9711200}}.

\bibitem[Polyakov(1998)]{Polyakov:1997tj}
A.~M. Polyakov, ``String theory and quark confinement'', {\em Nucl.\ Phys.\
  Proc.\ Suppl.} {\bf 68} (1998) 1--8,
\href{http://xxx.lanl.gov/abs/hep-th/9711002}{{\tt hep-th/9711002}}.

\bibitem[Polchinski(1995)]{polc1}
J.~Polchinski, ``Dirichlet-branes and {Ramond-Ramond} charges'', {\em Phys.\
  Rev.\ Lett.} {\bf 75} (1995) 4724,
  \href{http://xxx.lanl.gov/abs/hep-th/9510017}{{\tt hep-th/9510017}}.

\bibitem[Kuti(2006)]{Kuti:2005xg}
J.~Kuti, ``{Lattice QCD and string theory}'', {\em PoS} {\bf LAT2005} (2006)
  001,
\href{http://xxx.lanl.gov/abs/hep-lat/0511023}{{\tt hep-lat/0511023}}.

\bibitem[Muller and Nagle(2006)]{Muller:2006ee}
B.~Muller and J.~L. Nagle, ``Results from the relativistic heavy ion
  collider'', {\em Ann.\ Rev.\ Nucl.\ Part.\ Sci.} {\bf 56} (2006) 93--135,
\href{http://xxx.lanl.gov/abs/nucl-th/0602029}{{\tt nucl-th/0602029}}.

\bibitem[Beisert(2005)]{Beisert:2004ry}
N.~Beisert, ``{The dilatation operator of $N=4$ super Yang-Mills theory and
  integrability}'', {\em Phys.\ Rev.} {\bf 405} (2005) 1--202,
\href{http://xxx.lanl.gov/abs/hep-th/0407277}{{\tt hep-th/0407277}}.

\bibitem['t~Hooft(1974)]{'tHooft:1973jz}
G.~'t~Hooft, ``A planar diagram theory for strong interactions'', {\em Nucl.\
  Phys.} {\bf B72} (1974)
461.

\bibitem[Gopakumar(2004)]{Gopakumar:2003ns}
R.~Gopakumar, ``{From free fields to AdS}'', {\em Phys.\ Rev.} {\bf D70} (2004)
  025009,
\href{http://xxx.lanl.gov/abs/hep-th/0308184}{{\tt hep-th/0308184}}.

\bibitem[Wilczek(2004)]{Wilczek:2004im}
F.~Wilczek, ``Diquarks as inspiration and as objects'',
\href{http://xxx.lanl.gov/abs/hep-ph/0409168}{{\tt hep-ph/0409168}}.

\bibitem[Eidelman~et al.(2004)]{PDBook}
S.~Eidelman~et al., ``{Review of Particle Physics}'', {\em {Physics Letters B}}
  {\bf 592} (2004) 1.

\bibitem[Fradkin and Tseytlin(1985)]{Fradkin:1985qd}
E.~S. Fradkin and A.~A. Tseytlin, ``Nonlinear electrodynamics from quantized
  strings'', {\em Phys.\ Lett.} {\bf B163} (1985)
123.

\bibitem[Horowitz and Strominger(1991)]{horo4}
G.~T. Horowitz and A.~Strominger, ``Black strings and $p$-branes'', {\em Nucl.\
  Phys.} {\bf B360} (1991) 197--209.

\bibitem[Gubser et~al.(1998)Gubser, Klebanov, and Polyakov]{Gubser:1998bc}
S.~S. Gubser, I.~R. Klebanov, and A.~M. Polyakov, ``Gauge theory correlators
  from non-critical string theory'', {\em Phys.\ Lett.} {\bf B428} (1998)
  105--114,
\href{http://xxx.lanl.gov/abs/hep-th/9802109}{{\tt hep-th/9802109}}.

\bibitem[Witten(1998)]{Witten:1998qj}
E.~Witten, ``{Anti-de Sitter} space and holography'', {\em Adv.\ Theor.\ Math.\
  Phys.} {\bf 2} (1998) 253--291,
\href{http://xxx.lanl.gov/abs/hep-th/9802150}{{\tt hep-th/9802150}}.

\bibitem[Freedman et~al.(1999)Freedman, Mathur, Matusis, and
  Rastelli]{Freedman:1998tz}
D.~Z. Freedman, S.~D. Mathur, A.~Matusis, and L.~Rastelli, ``{Correlation
  functions in the CFT($d$)/AdS($d+1$) correspondence}'', {\em Nucl.\ Phys.}
  {\bf B546} (1999) 96--118,
\href{http://xxx.lanl.gov/abs/hep-th/9804058}{{\tt hep-th/9804058}}.

\bibitem[Aharony et~al.(2000)Aharony, Gubser, Maldacena, Ooguri, and
  Oz]{Aharony:1999ti}
O.~Aharony, S.~S. Gubser, J.~M. Maldacena, H.~Ooguri, and Y.~Oz, ``{Large-$N$
  field theories, string theory and gravity}'', {\em Phys.\ Rev.} {\bf 323}
  (2000) 183--386,
\href{http://xxx.lanl.gov/abs/hep-th/9905111}{{\tt hep-th/9905111}}.

\bibitem[Son and Starinets(2002)]{Son:2002sd}
D.~T. Son and A.~O. Starinets, ``{Minkowski-space correlators in AdS/CFT
  correspondence: recipe and applications}'', {\em JHEP\,} {\bf 09} (2002) 042,
\href{http://xxx.lanl.gov/abs/hep-th/0205051}{{\tt hep-th/0205051}}.

\bibitem[Csaki et~al.(1999)Csaki, Ooguri, Oz, and Terning]{Csaki:1998qr}
C.~Csaki, H.~Ooguri, Y.~Oz, and J.~Terning, ``Glueball mass spectrum from
  supergravity'', {\em JHEP\,} {\bf 01} (1999) 017,
\href{http://xxx.lanl.gov/abs/hep-th/9806021}{{\tt hep-th/9806021}}.

\bibitem[de~Mello~Koch et~al.(1998)de~Mello~Koch, Jevicki, Mihailescu, and
  Nunes]{deMelloKoch:1998qs}
R.~de~Mello~Koch, A.~Jevicki, M.~Mihailescu, and J.~P. Nunes, ``Evaluation of
  glueball masses from supergravity'', {\em Phys.\ Rev.} {\bf D58} (1998)
  105009,
\href{http://xxx.lanl.gov/abs/hep-th/9806125}{{\tt hep-th/9806125}}.

\bibitem[Maldacena(1998)]{Maldacena:1998im}
J.~M. Maldacena, ``{Wilson loops in large-$N$ field theories}'', {\em Phys.\
  Rev.\ Lett.} {\bf 80} (1998) 4859--4862,
\href{http://xxx.lanl.gov/abs/hep-th/9803002}{{\tt hep-th/9803002}}.

\bibitem[Kinar et~al.(2000)Kinar, Schreiber, and Sonnenschein]{Kinar:1998vq}
Y.~Kinar, E.~Schreiber, and J.~Sonnenschein, ``{$Q \bar{Q}$ potential from
  strings in curved spacetime -- classical results}'', {\em Nucl.\ Phys.} {\bf
  B566} (2000) 103--125,
\href{http://xxx.lanl.gov/abs/hep-th/9811192}{{\tt hep-th/9811192}}.

\bibitem[Polchinski and Strassler(2002)]{Polchinski:2001tt}
J.~Polchinski and M.~J. Strassler, ``Hard scattering and gauge/string
  duality'', {\em Phys.\ Rev.\ Lett.} {\bf 88} (2002) 031601,
\href{http://xxx.lanl.gov/abs/hep-th/0109174}{{\tt hep-th/0109174}}.

\bibitem[Erlich et~al.(2005)Erlich, Katz, Son, and Stephanov]{Erlich:2005qh}
J.~Erlich, E.~Katz, D.~T. Son, and M.~A. Stephanov, ``{QCD and a holographic
  model of hadrons}'', {\em Phys.\ Rev.\ Lett.} {\bf 95} (2005) 261602,
\href{http://xxx.lanl.gov/abs/hep-ph/0501128}{{\tt hep-ph/0501128}}.

\bibitem[de~Teramond and Brodsky(2005)]{deTeramond:2005su}
G.~F. de~Teramond and S.~J. Brodsky, ``{The hadronic spectrum of a holographic
  dual of QCD}'', {\em Phys.\ Rev.\ Lett.} {\bf 94} (2005) 201601,
\href{http://xxx.lanl.gov/abs/hep-th/0501022}{{\tt hep-th/0501022}}.

\bibitem[Gursoy and Kiritsis(2007)]{Gursoy:2007cb}
U.~Gursoy and E.~Kiritsis, ``Exploring improved holographic theories for {QCD:
  Part I}'',
\href{http://xxx.lanl.gov/abs/0707.1324}{{\tt arXiv:0707.1324 [hep-th]}}.

\bibitem[Gursoy et~al.(2007)Gursoy, Kiritsis, and Nitti]{Gursoy:2007er}
U.~Gursoy, E.~Kiritsis, and F.~Nitti, ``Exploring improved holographic theories
  for {QCD: Part II}'',
\href{http://xxx.lanl.gov/abs/0707.1349}{{\tt arXiv:0707.1349 [hep-th]}}.

\bibitem[Witten(1998)]{Witten:1998zw}
E.~Witten, ``{Anti-de Sitter} space, thermal phase transition, and confinement
  in gauge theories'', {\em Adv.\ Theor.\ Math.\ Phys.} {\bf 2} (1998) 505,
\href{http://xxx.lanl.gov/abs/hep-th/9803131}{{\tt hep-th/9803131}}.

\bibitem[Klebanov and Tseytlin(1996)]{Klebanov:1996un}
I.~R. Klebanov and A.~A. Tseytlin, ``{Entropy of Near-Extremal Black
  $p$-branes}'', {\em Nucl.\ Phys.} {\bf B475} (1996) 164--178,
\href{http://xxx.lanl.gov/abs/hep-th/9604089}{{\tt hep-th/9604089}}.

\bibitem[Itzhaki et~al.(1998)Itzhaki, Maldacena, Sonnenschein, and
  Yankielowicz]{Itzhaki:1998dd}
N.~Itzhaki, J.~M. Maldacena, J.~Sonnenschein, and S.~Yankielowicz,
  ``{Supergravity and the large-$N$ limit of theories with sixteen
  supercharges}'', {\em Phys.\ Rev.} {\bf D58} (1998) 046004,
\href{http://xxx.lanl.gov/abs/hep-th/9802042}{{\tt hep-th/9802042}}.

\bibitem[Polchinski and Strassler(2000)]{Polchinski:2000uf}
J.~Polchinski and M.~J. Strassler, ``The string dual of a confining
  four-dimensional gauge theory'',
\href{http://xxx.lanl.gov/abs/hep-th/0003136}{{\tt hep-th/0003136}}.

\bibitem[Klebanov and Strassler(2000)]{Klebanov:2000hb}
I.~R. Klebanov and M.~J. Strassler, ``Supergravity and a confining gauge
  theory: Duality cascades and {$\chi$SB-resolution} of naked singularities'',
  {\em JHEP\,} {\bf 08} (2000) 052,
\href{http://xxx.lanl.gov/abs/hep-th/0007191}{{\tt hep-th/0007191}}.

\bibitem[Maldacena and {Nu\~{n}ez}(2001)]{Maldacena:2000yy}
J.~M. Maldacena and C.~{Nu\~{n}ez}, ``Towards the large {$N$} limit of pure
  {$N=1$} super {Yang Mills}'', {\em Phys.\ Rev.\ Lett.} {\bf 86} (2001)
  588--591,
\href{http://xxx.lanl.gov/abs/hep-th/0008001}{{\tt hep-th/0008001}}.

\bibitem[Aharony(2002)]{Aharony:2002up}
O.~Aharony, ``{The non-AdS/non-CFT correspondence, or three different paths to
  QCD}'',
\href{http://xxx.lanl.gov/abs/hep-th/0212193}{{\tt hep-th/0212193}}.

\bibitem[Karch and Katz(2002)]{Karch:2002sh}
A.~Karch and E.~Katz, ``{Adding flavor to AdS/CFT}'', {\em JHEP\,} {\bf 06}
  (2002) 043,
\href{http://xxx.lanl.gov/abs/hep-th/0205236}{{\tt hep-th/0205236}}.

\bibitem[Sakai and Sonnenschein(2003)]{Sakai:2003wu}
T.~Sakai and J.~Sonnenschein, ``Probing flavored mesons of confining gauge
  theories by supergravity'', {\em JHEP\,} {\bf 09} (2003) 047,
\href{http://xxx.lanl.gov/abs/hep-th/0305049}{{\tt hep-th/0305049}}.

\bibitem[Sakai and Sugimoto(2005)]{Sakai:2004cn}
T.~Sakai and S.~Sugimoto, ``Low energy hadron physics in holographic {QCD}'',
  {\em Prog.\ Theor.\ Phys.} {\bf 113} (2005) 843--882,
\href{http://xxx.lanl.gov/abs/hep-th/0412141}{{\tt hep-th/0412141}}.

\bibitem[Babington et~al.(2004)Babington, Erdmenger, Evans, Guralnik, and
  Kirsch]{Babington:2003vm}
J.~Babington, J.~Erdmenger, N.~J. Evans, Z.~Guralnik, and I.~Kirsch, ``Chiral
  symmetry breaking and pions in non-supersymmetric gauge/gravity duals'', {\em
  Phys.\ Rev.} {\bf D69} (2004) 066007,
\href{http://xxx.lanl.gov/abs/hep-th/0306018}{{\tt hep-th/0306018}}.

\bibitem[Kruczenski et~al.(2004)Kruczenski, Mateos, Myers, and
  Winters]{Kruczenski:2003uq}
M.~Kruczenski, D.~Mateos, R.~C. Myers, and D.~J. Winters, ``{Towards a
  holographic dual of large-$N_c$ QCD}'', {\em JHEP\,} {\bf 05} (2004) 041,
\href{http://xxx.lanl.gov/abs/hep-th/0311270}{{\tt hep-th/0311270}}.

\bibitem[Casero et~al.(2007)Casero, Kiritsis, and Paredes]{Casero:2007ae}
R.~Casero, E.~Kiritsis, and A.~Paredes, ``Chiral symmetry breaking as open
  string tachyon condensation'', {\em Nucl.\ Phys.} {\bf B787} (2007) 98--134,
\href{http://xxx.lanl.gov/abs/hep-th/0702155}{{\tt hep-th/0702155}}.

\bibitem[Kruczenski et~al.(2005)Kruczenski, Zayas, Sonnenschein, and
  Vaman]{Kruczenski:2004me}
M.~Kruczenski, L.~A.~P. Zayas, J.~Sonnenschein, and D.~Vaman, ``{Regge
  trajectories for mesons in the holographic dual of large-$N_c$ QCD}'', {\em
  JHEP\,} {\bf 06} (2005) 046,
\href{http://xxx.lanl.gov/abs/hep-th/0410035}{{\tt hep-th/0410035}}.

\bibitem[{Sj\"ostrand}(1982)]{Sjostrand:1982fn}
T.~{Sj\"ostrand}, ``{The Lund Monte Carlo for jet fragmentation}'', {\em Comp.\
  Phys.\ Commun.} {\bf 27} (1982)
243.

\bibitem[Andersson et~al.(1983)Andersson, Gustafson, Ingelman, and
  {Sj\"ostrand}]{Andersson:1983ia}
B.~Andersson, G.~Gustafson, G.~Ingelman, and T.~{Sj\"ostrand}, ``Parton
  fragmentation and string dynamics'', {\em Phys.\ Rep.} {\bf 97} (1983)
31.

\bibitem[Casher et~al.(1979)Casher, Neuberger, and Nussinov]{Casher:1978wy}
A.~Casher, H.~Neuberger, and S.~Nussinov, ``Chromoelectric flux tube model of
  particle production'', {\em Phys.\ Rev.} {\bf D20} (1979)
179--188.

\bibitem[Gupta and Rosenzweig(1994)]{Gupta:1994tx}
K.~S. Gupta and C.~Rosenzweig, ``Semiclassical decay of excited string states
  on leading regge trajectories'', {\em Phys.\ Rev.} {\bf D50} (1994)
  3368--3376,
\href{http://xxx.lanl.gov/abs/hep-ph/9402263}{{\tt hep-ph/9402263}}.

\bibitem[Peeters et~al.(2006)Peeters, Sonnenschein, and
  Zamaklar]{Peeters:2005fq}
K.~Peeters, J.~Sonnenschein, and M.~Zamaklar, ``Holographic decays of
  large-spin mesons'', {\em JHEP\,} {\bf 02} (2006) 009,
\href{http://xxx.lanl.gov/abs/hep-th/0511044}{{\tt hep-th/0511044}}.

\bibitem[Aharony et~al.(2007)Aharony, Sonnenschein, and
  Yankielowicz]{Aharony:2006da}
O.~Aharony, J.~Sonnenschein, and S.~Yankielowicz, ``A holographic model of
  deconfinement and chiral symmetry restoration'', {\em Ann.\ Phys.} {\bf 322}
  (2007) 1420--1443,
\href{http://xxx.lanl.gov/abs/hep-th/0604161}{{\tt hep-th/0604161}}.

\bibitem[Peeters et~al.(2006)Peeters, Sonnenschein, and
  Zamaklar]{Peeters:2006iu}
K.~Peeters, J.~Sonnenschein, and M.~Zamaklar, ``Holographic melting and related
  properties of mesons in a quark gluon plasma'', {\em Phys.\ Rev.} {\bf D74}
  (2006) 106008,
\href{http://xxx.lanl.gov/abs/hep-th/0606195}{{\tt hep-th/0606195}}.

\bibitem[Karsch(2000)]{Karsch:1999vy}
F.~Karsch, ``{Lattice QCD at finite temperature and density}'', {\em Nucl.\
  Phys.\ Proc.\ Suppl.} {\bf 83} (2000) 14--23,
\href{http://xxx.lanl.gov/abs/hep-lat/9909006}{{\tt hep-lat/9909006}}.

\bibitem[Gottlieb et~al.(1997)]{Gottlieb:1996ae}
S.~A. Gottlieb {\em et~al.}, ``{Thermodynamics of lattice QCD with two light
  quark flavours on a $16^3 \times 8$ lattice. II}'', {\em Phys.\ Rev.} {\bf
  D55} (1997) 6852--6860,
\href{http://xxx.lanl.gov/abs/hep-lat/9612020}{{\tt hep-lat/9612020}}.

\bibitem[Shuryak(2005)]{Shuryak:2004cy}
E.~V. Shuryak, ``{What RHIC experiments and theory tell us about properties of
  quark-gluon plasma?}'', {\em Nucl.\ Phys.} {\bf A750} (2005) 64--83,
\href{http://xxx.lanl.gov/abs/hep-ph/0405066}{{\tt hep-ph/0405066}}.

\bibitem[Teaney(2003)]{Teaney:2003kp}
D.~Teaney, ``{Effect of shear viscosity on spectra, elliptic flow, and Hanbury
  Brown-Twiss radii}'', {\em Phys.\ Rev.} {\bf C68} (2003) 034913,
\href{http://xxx.lanl.gov/abs/nucl-th/0301099}{{\tt nucl-th/0301099}}.

\bibitem[Son and Starinets(2007)]{Son:2007vk}
D.~T. Son and A.~O. Starinets, ``Viscosity, black holes, and quantum field
  theory'',
\href{http://xxx.lanl.gov/abs/0704.0240}{{\tt arXiv:0704.0240 [hep-th]}}.

\bibitem[Herzog et~al.(2006)Herzog, Karch, Kovtun, Kozcaz, and
  Yaffe]{Herzog:2006gh}
C.~P. Herzog, A.~Karch, P.~Kovtun, C.~Kozcaz, and L.~G. Yaffe, ``Energy loss of
  a heavy quark moving through {$N\!=\!4$} supersymmetric {Yang-Mills}
  plasma'', {\em JHEP\,} {\bf 07} (2006) 013,
\href{http://xxx.lanl.gov/abs/hep-th/0605158}{{\tt hep-th/0605158}}.

\bibitem[Gubser(2006)]{Gubser:2006bz}
S.~S. Gubser, ``{Drag force in AdS/CFT}'', {\em Phys.\ Rev.} {\bf D74} (2006)
  126005,
\href{http://xxx.lanl.gov/abs/hep-th/0605182}{{\tt hep-th/0605182}}.

\bibitem[Burikham and Li(2007)]{Burikham:2007kp}
P.~Burikham and J.~Li, ``Aspects of the screening length and drag force in two
  alternative gravity duals of the quark-gluon plasma'', {\em JHEP\,} {\bf 03}
  (2007) 067,
\href{http://xxx.lanl.gov/abs/hep-ph/0701259}{{\tt hep-ph/0701259}}.

\bibitem[Buchel(2006)]{Buchel:2006bv}
A.~Buchel, ``On jet quenching parameters in strongly coupled non- conformal
  gauge theories'', {\em Phys. Rev.} {\bf D74} (2006) 046006,
\href{http://xxx.lanl.gov/abs/hep-th/0605178}{{\tt hep-th/0605178}}.

\bibitem[Armesto et~al.(2006)Armesto, Edelstein, and Mas]{Armesto:2006zv}
N.~Armesto, J.~D. Edelstein, and J.~Mas, ``{Jet quenching at finite 't~Hooft
  coupling and chemical potential from AdS/CFT}'', {\em JHEP\,} {\bf 09} (2006)
  039,
\href{http://xxx.lanl.gov/abs/hep-ph/0606245}{{\tt hep-ph/0606245}}.

\bibitem[Kaczmarek and Zantow(2005)]{Kaczmarek:2005ui}
O.~Kaczmarek and F.~Zantow, ``{Static quark anti-quark interactions in zero and
  finite temperature QCD. I: Heavy quark free energies, running coupling and
  quarkonium binding}'', {\em Phys.\ Rev.} {\bf D71} (2005) 114510,
\href{http://xxx.lanl.gov/abs/hep-lat/0503017}{{\tt hep-lat/0503017}}.

\bibitem[Liu et~al.(2007)Liu, Rajagopal, and Wiedemann]{Liu:2006nn}
H.~Liu, K.~Rajagopal, and U.~A. Wiedemann, ``An {AdS/CFT} calculation of
  screening in a hot wind'', {\em Phys.\ Rev.\ Lett.} {\bf 98} (2007) 182301,
\href{http://xxx.lanl.gov/abs/hep-ph/0607062}{{\tt hep-ph/0607062}}.

\bibitem[Chernicoff et~al.(2006)Chernicoff, Garcia, and
  Guijosa]{Chernicoff:2006hi}
M.~Chernicoff, J.~A. Garcia, and A.~Guijosa, ``{The energy of a moving
  quark-antiquark pair in an {$N\!=\!4$} SYM plasma}'', {\em JHEP\,} {\bf 09}
  (2006) 068,
\href{http://xxx.lanl.gov/abs/hep-th/0607089}{{\tt hep-th/0607089}}.

\end{thebibliography}

\begingroup\raggedright\endgroup

\end{document}